\documentclass[fleqn,usenatbib]{mnras}

\usepackage{newtxtext,newtxmath}

\usepackage[T1]{fontenc}

\DeclareRobustCommand{\VAN}[3]{#2}
\let\VANthebibliography\thebibliography
\def\thebibliography{\DeclareRobustCommand{\VAN}[3]{##3}\VANthebibliography}

\usepackage{graphicx}	
\usepackage{amsmath}	

\title[Soft X-ray Scattered Emission in obscured AGN]{Ray-tracing simulations of the  Soft X-ray Scattered Emission in obscured Active Galactic Nuclei}

\author[McKaig et al.]{
Jeffrey McKaig$^{1}$\thanks{E-mail: jmckaig@gmu.edu},
Claudio Ricci$^{2,3,1}$,
St\'ephane Paltani$^{4}$,
K. K. Gupta$^{2,5}$,
Nicholas P. Abel$^{6}$,
Y. Ueda$^{7}$
\\
$^{1}$George Mason University, Department of Physics and Astronomy, MS3F3, 4400 University Drive, Fairfax, VA 22030, USA\\
$^{2}$Instituto de Estudios Astrof\'isicos, Facultad de Ingenier\'ia y Ciencias, Universidad Diego Portales, Av. Ej\'ercito Libertador 441, Santiago, Chile\\
$^{3}$Kavli Institute for Astronomy and Astrophysics, Peking University, Beijing 100871, China\\
$^{4}$Department of Astronomy, University of Geneva, 1205 Versoix, Switzerland\\
$^5$STAR Institute, Quartier Agora - All\'ee du six Ao\^ut, 19c B-4000 Li\`ege, Belgium\\
$^{6}$College of Applied Science, University of Cincinnati, Cincinnati, OH, 45206, USA\\
$^{7}$Department of Astronomy, Kyoto University, Kyoto 606-8502, Japan
}
\date{Accepted 2023 September 16. Received 2023 September 12; in original form 2023 May 9}

\pubyear{2015}

\begin{document}
\label{firstpage}
\pagerange{\pageref{firstpage}--\pageref{lastpage}}
\maketitle

\begin{abstract}
Most Active Galactic Nuclei (AGN) in the local Universe are obscured. In these obscured AGN an excess is usually observed in the soft X-rays below $\sim2$\,keV above the absorbed X-ray continuum. This spectral component is associated with the scattering of X-ray photons off free electrons in the Narrow Line Region (NLR), and/or to photoionised lines. Recent studies have found that in highly obscured AGN this component has lower flux relative to the primary X-ray continuum than in less obscured AGN. This is measured by the scattering fraction, or $f_{\text{scatt}}$, which is the ratio of the scattered flux to the continuum. Here, we use the ray-tracing platform \textsc{RefleX} to perform simulations of scattered X-ray radiation to test two possible explanations for this phenomenon: (1) sources with lower $f_{\text{scatt}}$ are viewed at higher inclinations or (2) low $f_{\text{scatt}}$ sources are characterised by larger covering factors. We consider a conical NLR of free electrons, while allowing the column density and opening angle (and hence covering factor) to vary. We also consider electron densities inferred from observations, and from simulations carried out with the spectral synthesis code {\sc Cloudy}. Our simulations show $f_{\text{scatt}}$ is expected to be related to both the inclination angle and covering factor of the torus; however, the observed negative correlation between $f_{\text{scatt}}$ and $N_{\rm H}$ can only be explained by a positive relation between the column density and the covering factor of the obscuring material. Additional contributions to $f_{\text{scatt}}$ can come from unresolved photoionised lines and ionised outflowing gas.    
\end{abstract}

\begin{keywords}
atomic processes – galaxies: Seyfert - galaxies: active - X-rays: general - X-rays: galaxies
\end{keywords}



\section{Introduction}\label{sect:intro}

    It is now widely accepted that the high luminosity and wide ranging continuum seen in active galactic nuclei (AGN) is the result of mass accretion onto a Supermassive Black Hole (SMBH) at the centre of the host galaxy. In-flowing material is expected to form an accretion disk around the SMBH (e.g., \citealp{1973A&A....24..337S}). This accretion flow produces copious optical / UV photons, which can then be up-scattered into the X-ray band by a corona of hot electrons \citep{1991ApJ...380L..51H, 1993ApJ...413..507H}. These X-ray photons are absorbed and reprocessed by various distributions of gas and dust surrounding the SMBH, such as the broad line region (BLR) and the dusty torus in the plane of the accretion disk, as well as the narrow line region (NLR), which is expected to be perpendicular to the accretion disk (e.g., \citealp{1995PASP..107..803U}). The fact that the system is axially symmetric is thought to be one of the main reasons we observe different classes of AGN (i.e., type\,1s and type\,2s; e.g., \citealp{doi:10.1146/annurev.aa.31.090193.002353, doi:10.1146/annurev-astro-082214-122302, 2017NatAs...1..679R}).
    
    The primary X-ray emission component, observed in AGN as a powerlaw [$F(E)\propto E^{-\Gamma}$; \citealp{1993ARA&A..31..717M, 2008A&A...485..417D, 2009A&A...505..417B}], is thought to be produced in the X-ray corona where photons created in the accretion disk are Compton up-scattered into X-ray energies via optically thin, hot electrons surrounding the accretion disk. The powerlaw is characterised by a photon index of $\Gamma\sim1.8-1.9$ \citep{Nandra7469,10.1111/j.1365-2966.2011.18207.x, 2017ApJS..233...17R}, and shows an exponential cutoff at a few hundred keV \citep{2008A&A...485..417D,10.1093/mnras/sty1879,2020ApJ...905...41B}. X-ray emission in AGN is accompanied by a reflection component, which is produced by X-ray photons reprocessed by the circumnuclear material. This is typically observed as an Fe K$\alpha$ line at 6.4\,keV \citep{2000PASP..112.1145F, 2004ApJ...604...63Y, 2010ApJS..187..581S, 2011ApJ...727...19F, 2014MNRAS.441.3622R} and a reflection hump (known as the ``Compton Hump'') at $\sim$30\,keV \citep{1988ApJ...335...57L, 1994MNRAS.267..743G, 1994ApJ...420L..57K}. X-ray absorption also has a significant effect on the observed spectra of AGN. The main components driving absorption in the X-rays are: i) photoelectric absorption which, for soft X-rays ($<10$\,keV) can play a significant role at column densities as low as $\log{(N_{\text{H}}\big/\text{\,cm}^{-2})} \approx 21-22$ and ii) Compton scattering from bound electrons at column densities of $\log{(N_{\text{H}}\big/\text{\,cm}^{-2})} \approx 23.5$. 
    
    X-ray surveys have been increasingly used to detect and identify AGN, especially obscured AGN, for which traditional optical diagnostics can be uncertain (e.g., \citealp{2015A&ARv..23....1B,2017ApJ...850...74K,2017ApJS..233...17R,2018ARA&A..56..625H}). This is due to the fact that hard X-rays ($E > 10\,$keV) can penetrate high column densities of material as well as avoid contamination from the host galaxy emission, which allows the detection and study of highly obscured, even Compton thick AGN [$\log{(N_{\text{H}}\big/\text{\,cm}^{-2})} \geq 24$; \citealp{10.1046/j.1365-8711.2000.03721.x,2007ApJ...664L..79U,10.1111/j.1365-2966.2010.17902.x,10.1111/j.1365-2966.2012.20908.x,2015ApJ...815L..13R, 10.1093/mnras/stw1764, 2017ApJS..233...17R,Hikitani_2018}]. The obscured phase could represent an important stage in the life of AGN, with this period being associated to a significant fraction of black hole growth. In some cases this heavily obscured phase is thought to be onset from a merger event (e.g., \citealp{2018MNRAS.478.3056B,Ricci:2017aa,Ricci:2021oz,Yamada:2021vz}). Thus, X-ray surveys are an important tool for studying SMBH and galaxy evolution.
    
    In all obscured AGN, an excess at $E\lesssim 2-3$\,keV over the absorbed continuum is observed. This soft excess is very different from that usually observed in unobscured AGN, which is likely produced very close to the SMBH (e.g., \citealp{Gierlinski:2004wb,Crummy:2006wh,2012MNRAS.420.1848D,Boissay:2016nw,Petrucci:2018sz,Petrucci:2020ee,Gliozzi:2020zl}).
    The soft X-ray excess in obscured AGN has been interpreted as a combination of scattered radiation off electrons around the AGN \citep{1997ApJ...488..164T, 2006A&A...446..459C,2007ApJ...664L..79U,2017ApJS..233...17R} and photoionised lines \citep{2003ApJ...596..114S, 2006A&A...448..499B, 2007MNRAS.374.1290G}. The fact that photoionised gas can be traced with the [O\,{\sc iii}] $\lambda 5007$ emission out to hundreds of parsecs, along with the domination of photoionised lines, points to the soft X-ray excess being associated with gas in the NLR \citep{2000ApJ...543L.115S,2006A&A...448..499B, 2007MNRAS.374.1290G, 10.1093/mnras/stab839}. The amount of scattered radiation seen over the absorbed continuum can be quantified by the scattering fraction $f_{\text{scatt}}$ \citep{1997ApJ...488..164T, 2006A&A...446..459C,2007ApJ...664L..79U, 2009ApJ...690.1322W, 2016ApJ...831...37K, 2020ApJ...897..107Y}, which is the ratio of the flux of scattered radiation to the flux of the primary continuum. 
    
    Numerous studies have looked at the amount of scattered X-ray radiation in obscured AGN. For example, studying hard X-ray selected AGN from the \textit{Swift}/BAT all-sky-survey, \cite{2007ApJ...664L..79U} reported two AGN with unusually low values of $f_{\text{scatt}}$ ($<0.5\%$). They explained the low value as being due to the fact that these AGN are buried in a geometrically-thick torus, or to the absence of gas available for scattering. \cite{2012ApJ...754...45I} used a sample of 135 non-Blazar \textit{Swift}/BAT AGN that included 13 sources with $f_{\text{scatt}} < 0.5\%$ to study their infrared properties. They found enhanced fluxes at 9\,$\mu$m in the averaged SED of these AGN with low $f_{\text{scatt}}$ relative to the type\,2 AGN in the sample, which could suggest that the host galaxy is undergoing a starburst phase. Furthermore, AGN with $f_{\mathrm{scatt}}<0.5\%$ tend to also have low [O\,{\sc iv}] $\lambda$25.89 / $L(12\,\mu\mathrm{m})$ ratio, giving a new method of identifying ``buried'' AGN \citep{2019ApJ...876...96Y}. However, \cite{2014MNRAS.438..647H} argued that this new class of AGN with low $f_{\text{scatt}}$ could possibly be caused by host-galaxy contamination along the line of sight, since many of these AGN are in edge-on galaxies. \cite{2015ApJ...815....1U} used a complete sample of AGN from the \textit{Swift}/BAT 9-month hard X-ray survey in order to study the correlation between the X-ray and [O\,{\sc iii}] $\lambda5007$ luminosities. They concluded that AGN with low $f_{\text{scatt}}$ also showed low [O\,{\sc iii}] $\lambda5007$ to X-ray luminosity ratios, suggesting that their SMBHs are buried in tori with small opening angles. A similar conclusion was drawn in \cite{10.1093/mnras/stab839} who, from a sample of hard X-ray selected obscured AGN, found a positive correlation between $f_{\text{scatt}}$ and the [O\,{\sc iii}] $\lambda5007$ to X-ray luminosity ratio. \cite{2015ApJ...815....1U} also noted that in their sample of AGN with low $f_{\text{scatt}}$, more than half are free from host galaxy absorption by matter in the galactic disk. Therefore, absorption from the host galaxy alone cannot be the reason for the low $f_{\text{scatt}}$ values.  
    
    More recently, \citeauthor{10.1093/mnras/stab839} (\citeyear{10.1093/mnras/stab839}; hereafter G21) found a significant negative correlation between $f_{\text{scatt}}$ and the observed column density ($N_{\text{H}}$; i.e., the line-of-sight column density through an obscuring torus) in a sample of 386 hard X-ray selected obscured AGN from the BASS\footnote{\url{www.bass-survey.com}} survey (\citealp{2017ApJ...850...74K,Koss:2022qi,2017ApJS..233...17R}; see Figure\,1 and Table\,1 in G21). G21 performed spectral simulations of dummy populations of AGN that indicate this relation is not due to an inherent degeneracy between $f_{\text{scatt}}$ and $N_{\text{H}}$. They were also able to reproduce this correlation in different bins of Eddington ratio, black hole mass, and X-ray luminosity, further confirming that this correlation does not depend on these physical properties of AGN. Similar to \cite{2010ApJ...711..144N}, G21 found a slight negative correlation between $f_\mathrm{scatt}$ and the Eddington ratio with no correlation with black hole mass or X-ray luminosity. G21 put forward two different explanations for the $f_{\mathrm{scatt}}-N_{\mathrm{H}}$ relation: (1) the dependence of the differential Compton cross-section on the inclination angle or (2) sources with low $f_{\text{scatt}}$ might tend to have larger covering factors ($f_{\text{cov}}$; fraction of the sky covered by the obscuring torus) as proposed by \cite{2007ApJ...664L..79U}, reducing the amount of scattered radiation. This would imply a relationship between the covering factor of a source and its column density (e.g., \citealp{2011ApJ...731...92R, 2011A&A...532A.102R, 2016ApJ...819..166M, 2018ApJ...853..146T, 2019A&A...626A..40P}; G21). The first explanation relies on the fact that the observed photons are Compton scattered off material in the NLR and therefore are subject to the Klein-Nishina (K-N) cross-section \citep{1986rpa..book.....R}. However, the photons responsible for the soft X-ray excess have energies $\leq$2\,keV and are therefore subject to the low energy approximation to the K-N form. This is the energy independent elastic Thomson cross-section with the differential relation $d\sigma/d\Omega \propto 1 + \cos^{2}(\theta)$, where $\theta$ is the scattering angle and $d\Omega = \cos{(\theta)d\theta d\phi}$. Thus, when these photons interact with the free electrons in the NLR, the electrons will behave like a mirror while re-directing the incoming photons to the line of sight. G21 then assumed a similar torus model for all the AGN in their sample while predicting the covering factor of different layers of the torus using statistical arguments from \citealp{2015ApJ...815L..13R}. Since the scattering fraction is expected to depend on the inclination angle, sources with higher column densities (which are being viewed at higher inclinations), will have lower values of $f_{\text{scatt}}$, consistent with the relationship between $f_{\text{scatt}}$ and $N_{\text{H}}$.
    
    Our goal in this paper is to use the ray-tracing simulation software \textsc{RefleX}\footnote{\url{https://www.astro.unige.ch/reflex/}} (\citealp{RefleX}, see also \citealp{Ricci:2023zk}) in order to study the origin of the scattered component with respect to Thomson scattered emission, and to test the two different explanations for the trend between $f_{\text{scatt}}$ and $N_{\text{H}}$: (1) a relation between the column density and the inclination angle to the AGN, and (2) a scaling relationship between the column density of a source and its covering factor, where sources with higher column densities have larger covering factors. To do this, we consider a NLR of various opening angles consisting of free electrons exposed to a powerlaw X-ray source to test the amount of scattered radiation at various inclination angles. We then use the inclination angle as a proxy for the torus column density to compare our relation to G21, discussing both the normalisation as well as the slope. Our model setup can be found in $\S$\ref{ModelSetup}, results in $\S$\ref{results}, discussion in $\S$\ref{Discussion}, and our conclusions in $\S$\ref{conclusions}.

            \begin{figure*}
                \centering
                \includegraphics[scale=0.5]{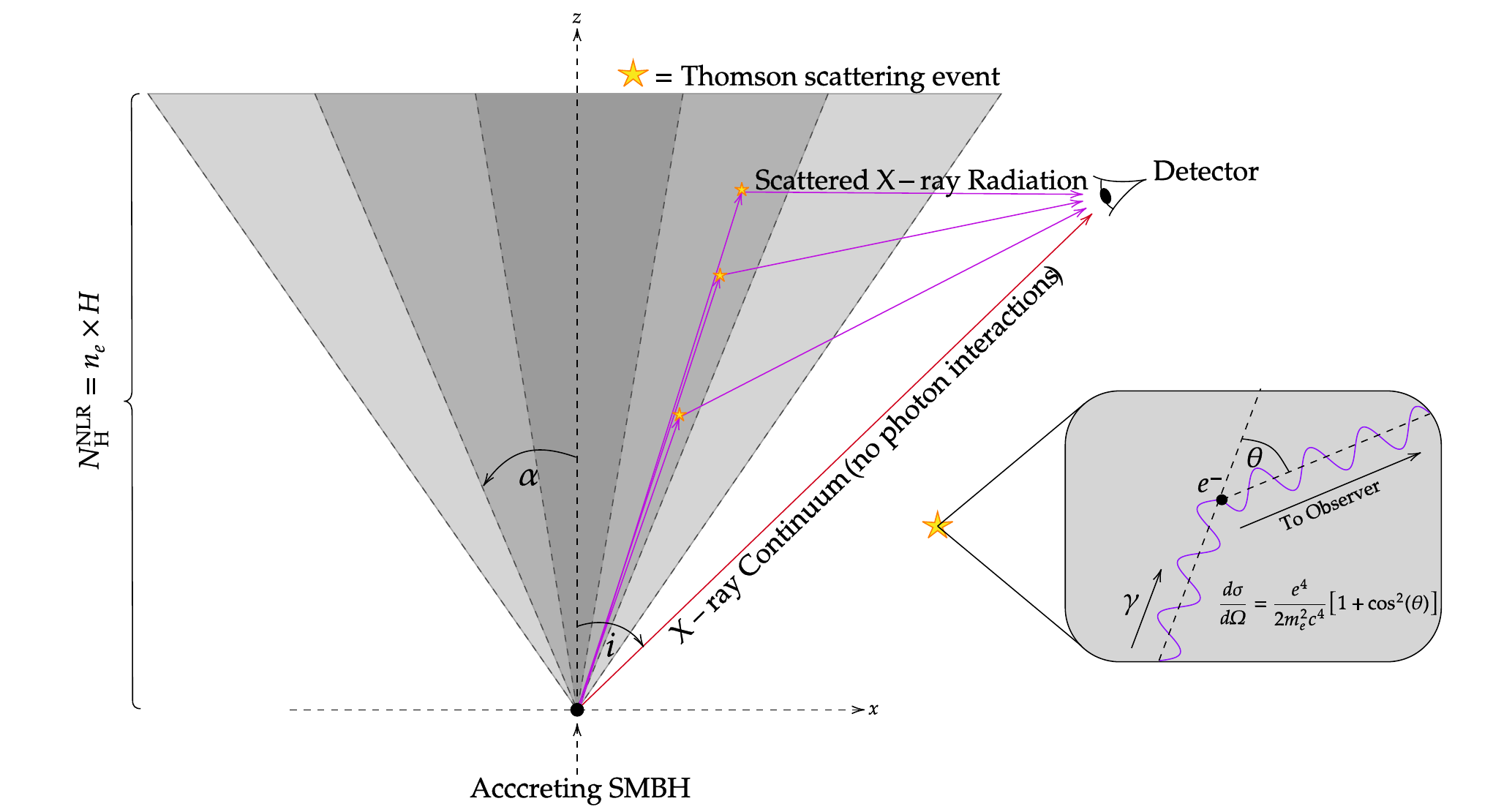}
                \caption{Cross section of the NLR geometry considered in our simulations and a diagram showing the Thomson scattering process with the incoming photon ($\gamma$) and electron $e^{-}$. Yellow stars with orange borders represent Thomson scattering events. The NLR traces a cone geometry as seen in optical kinematic and narrowband imaging. Each shade of grey shows a different half opening angle ($\alpha$; not all opening angles are shown). The inclination angle to the detector, $i$, is also shown. The scattering fraction $f_{\text{scatt}}$ is calculated by isolating the photons which undergo scattering off the free electrons in the NLR as well as the photons which undergo no interactions (the X-ray continuum), calculating the flux ratio of these two groups of photons in 5\,eV bins, and averaging the results. The column density is calculated by multiplying the electron density ($n_{e}$) by the height of the NLR ($H = 1000\,$pc).} 
                \label{fig1}
            \end{figure*}

    \section{Model Setup}\label{ModelSetup}
    
        \subsection{\textsc{RefleX}}\label{RefleX}
        
            In order to study the fraction of scattered X-ray radiation from AGN, we use the X-ray simulation platform \textsc{RefleX} \citep{RefleX}. \textsc{RefleX} uses ray-tracing to track photons with energies ranging from  0.1\,keV to 1\,MeV interacting with various distributions of gas and dust around an X-ray source. The X-ray source can output different spectral distributions of photons such as a powerlaw, Gaussian, or a blackbody. \textsc{RefleX} also allows the user to specify different combinations of geometries of gas such as a disk, torus, or a cone with user-specified sizes and densities. Material in these geometries can be neutral, ionised, molecular, or in the form of dust \citep{Ricci:2023zk}. Photons can undergo many different physical processes when interacting with the material, such as photo-ionisation, Compton scattering from free and bound electrons, and Rayleigh scattering. Therefore, with its ability to consider many different geometries and scattering processes, \textsc{RefleX} represents a very powerful tool to study how photons will interact with the free electrons in the NLR, while isolating the continuum (or non-interacting) photons.  
        
        \subsection{NLR Model: Geometry and Composition}\label{Geometry}
            
            \subsubsection{Geometry and incident radiation}
            
            We consider the traditional unification model where there torus collimates the radiation from the central accretion disk and forms a bi-cone shape for the NLR. This has been confirmed via kinematic (e.g., \citealp{2013ApJS..209....1F}) and optical narrowbnd imaging of [O\,\textsc{iii}] $\lambda$5007 and H$\alpha$+[N\,\textsc{ii}] $\lambda$6584 (e.g., \citealp{2018ApJ...868...14S}). The geometry for our simulated NLR can be seen in Figure \ref{fig1} where the various shades of grey shown represent different half opening angles ($\alpha$; not all shown) to be considered. The NLR extends from an X-ray point source located at the center of the system representing the accreting SMBH to a height of $H = 1\,$kpc. This range is generally observed in AGN (e.g., \citealp{1981ApJ...247..403H, 1987ApJ...321L..29M, 2013MNRAS.430.2327L, 2019MNRAS.489..855C}). The X-ray source emits a powerlaw spectrum incident on the NLR in the 0.3$-$200\,keV range with a photon index of $\Gamma = 1.8$ and exponential cutoff at 200\,keV, typical values for nearby AGN (e.g., \citealp{Nandra7469,10.1111/j.1365-2966.2011.18207.x, 2017ApJS..233...17R,10.1093/mnras/sty1879}). 
            
            \subsubsection{Expected NLR column density}\label{density} 
            
            Thomson scattering occurs when a soft X-ray photon interacts with a free electron (see Figure\,\ref{fig1}).  Thus, the free electron density distribution observed in the NLR may be an important parameter in our simulations. The properties of the NLR in AGN have been widely studied (e.g., \citealp{1989MNRAS.240..225T,2006A&A...448..499B, 2006A&A...459...55B, 2010A&A...516A...9D, 2013ApJS..209....1F, 2017MNRAS.469.2720G, Fabbiano_2018}). In particular, \cite{2006A&A...459...55B} used optical spectra from a sample of six Seyfert-1 galaxies to study the physical properties of their NLRs such as the electron density as a function of radius using a powerlaw:   
            \begin{equation}\label{eq1}
                n_{\text{e}}(r) = n_{\text{e,0}}\left(\frac{r}{r_{0}}\right)^{-\delta}
            \end{equation}
            where $n_{\text{e,0}}$ is the density of electrons at $r_{0}$, $r$ is the distance to the SMBH, and $\delta$ is a decay parameter. Table 9 in \cite{2006A&A...459...55B} shows the values of $n_{\mathrm{e,0}}$ and $\delta$ for all sources observed in their sample. These range from $3.1 \leq \log{(n_{\mathrm{e,0}}\big/\mathrm{cm}^{-3})} \leq 4.1$ and $-2.32 \leq \delta \leq -0.94$. The density drop off with distance generally observed in AGN (e.g., \citealp{2000ApJ...531..278K, 2001ApJ...546..205B, Walsh_2008, 2009A&A...500.1287S}) could be explained by the compression of the gas by the incident radiation pressure as shown by \cite{2014MNRAS.438..901S}. A discussion on the effect that this density distribution has on the fraction of scattered radiation can be found in Appendix \ref{AA}. There we show that the exact density distribution as a function of depth has little effect on the fraction of scattered radiation, and only the total column density of free electrons needs to be considered. Thus, our simulated NLR will have a constant density of free electrons\footnote{More accurately, \textsc{RefleX} allows for a population of fully ionised hydrogen and helium with the `\textsc{metallicity} 0 \textsc{temperature} 1' command. This is what was used here.} with distance from the SMBH. The column density of free electrons as a function of the density distribution can be found using the integral equation: 
            \begin{equation}\label{eq2}
                N_{\text{H}}^{\text{NLR}} = \int_{r_{0}}^{H} fn_{\mathrm{e}}(r)dr
            \end{equation}
             where $f$ is the filling factor of the gas, assumed to be unity, and $n_{\mathrm{e}}(r)$ is given by Equation\,\ref{eq1}. Considering different values of $f$ will change the NLR column density and thus may also be an important factor in our simulations. However, we will also consider different NLR column densities later in our calculations so we do not vary $f$. To find an estimate for the column density to use in our simulations, we average the values of $n_{\mathrm{e,0}}$ and $\delta$ from Table\,9 in \cite{2006A&A...459...55B} and use these in Equation\,\ref{eq2}. This gives a column density of $N_{\text{H}}^{\text{NLR}} = 10^{22.43}$\,cm$^{-2}$. To use this value in our simulations, we assume the column density is measured along the $z$ axis (see Figure\,\ref{fig1}) from the SMBH to the height of the NLR and calculate the corresponding density with $n_{\mathrm{e}} = N_{\mathrm{H}}^{\mathrm{NLR}} / H$.     
            
            \begin{figure}
                \centering
                \includegraphics[scale = 0.3]{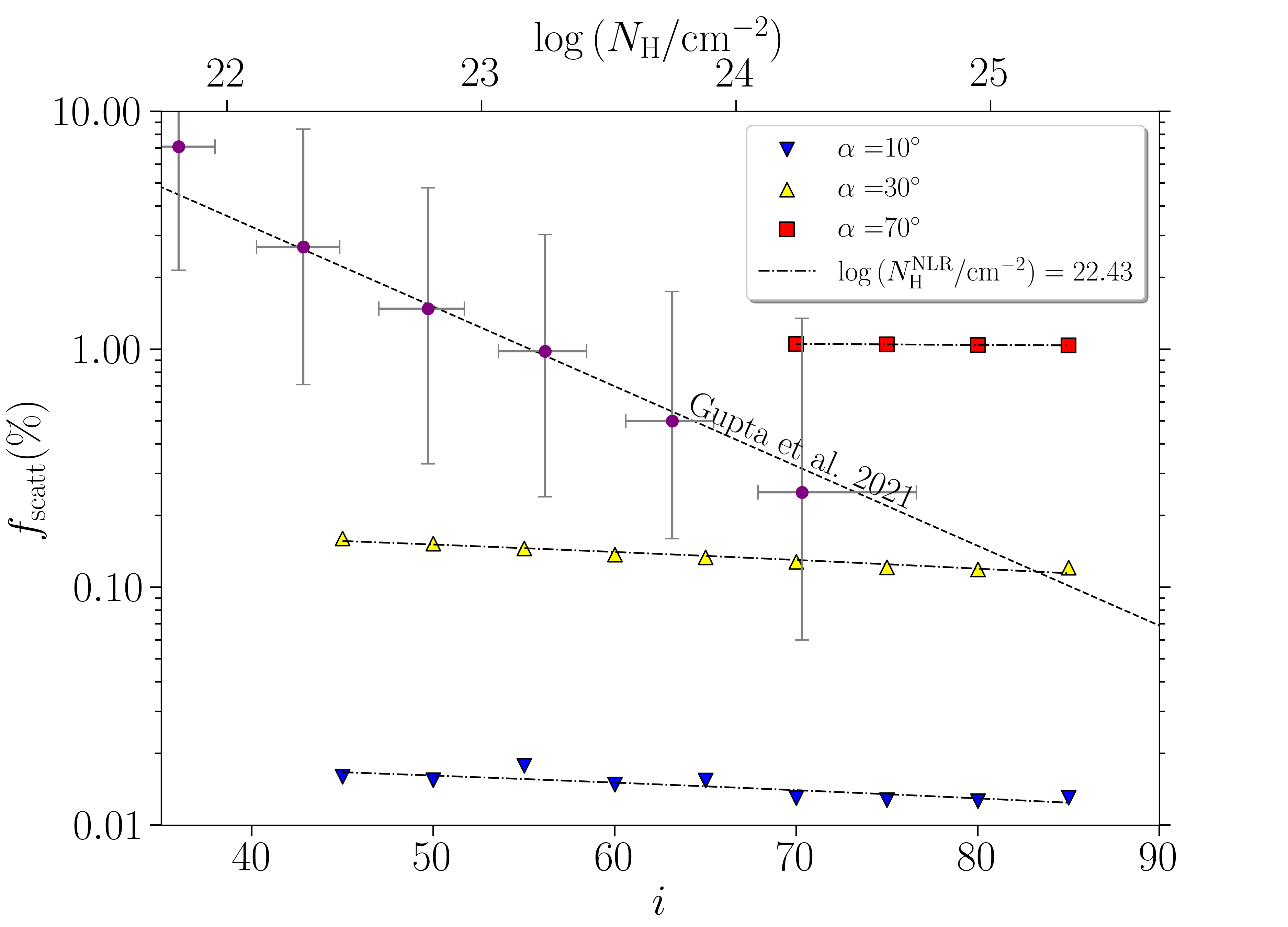}
                \caption{Fraction of scattered radiation as a function of inclination angle (bottom axis) and torus column density (top axis) for different NLR opening angles ($\alpha = 10^{\circ}, 30^{\circ}$, and $70^{\circ}$). Inclination angles are translated into column densities via Equation\,\ref{eq1p5} (see also \S\ref{slope} for details) based on the G21 model. Dashed-dotted lines show relations for the scattering fraction for the NLR column density ($N_{\text{H}}^{\text{NLR}}$) as calculated from averaging the sources from \protect\citeauthor{2006A&A...459...55B} (\citeyear{2006A&A...459...55B}$; N_{\text{H}}^{\text{NLR}} = 10^{22.43}\text{\,cm}^{-2}$). The dashed black line is the best fit trend to the data points (purple points with grey error bars) which are the mean $f_{\mathrm{scatt}}$ values per $N_{\mathrm{H}}$ bin found in G21.}
                \label{fig2}
            \end{figure}  
             
             In addition to the column density of free electrons as obtained from the method outlined above, we also use the spectral synthesis code {\sc Cloudy}\footnote{\url{https://www.nublado.org}} (version c17.02; \citealp{2017RMxAA..53..385F}) in order to infer a free electron column density in a 1-dimensional sphere of photoionized gas using an input AGN spectral energy distribution (SED). {\sc Cloudy} allows for the input of an SED, gas density, and composition, as well as various stopping criteria such as temperature, pressure, or cloud thickness. We use the \textsc{XSPEC} \citep{1996ASPC..101...17A} model \textsc{optxagnf} from  \cite{2012MNRAS.420.1848D} assuming a black hole mass of $M_{\text{BH}} = 10^{7.5}M_{\odot}$ and an Eddington ratio of $\log{(L/L_{\text{Edd}})} = -1.5$, values typical of nearby AGN (e.g., \citealp{2017ApJ...850...74K,Ricci2017}), as the incident AGN SED. The cloud is described by an input radius of 1\,pc, a stopping radius of 1000\,pc, a density law as seen in Equation\,\ref{eq1} with $n_{\mathrm{e, 0}}$ and $\delta$ taken as the average from the \cite{2006A&A...459...55B} sources. Additionally, the cloud has depleted ISM abundances with \textsc{Cloudy}'s defult ISM grain set included. A description of the other parameters used in the input SED, as well as other important {\sc Cloudy} parameters, is reported in Appendix \ref{AB}. The results of these {\sc Cloudy} simulations give a column density of $10^{22.38}\,\text{\,cm}^{-2}$ for free electrons from Hydrogen, the main contributor to the population of free electrons, at a depth of 1000\,pc. This is in good agreement with what was obtained by averaging the column densities from the \cite{2006A&A...459...55B} sources using Equation\,\ref{eq2}. We note that while \textsc{Cloudy} produces results for a 1-dimensional spherical geometry, different from the 3-dimensional ray-tracing implementation in \textsc{RefleX}, we calculate $N_{\mathrm{H}}^{\mathrm{NLR}}$ in a geometry-independent way and this can be compared to the Cloudy calculations. 
            
            \begin{figure}
                \centering
                \includegraphics[scale = 0.35]{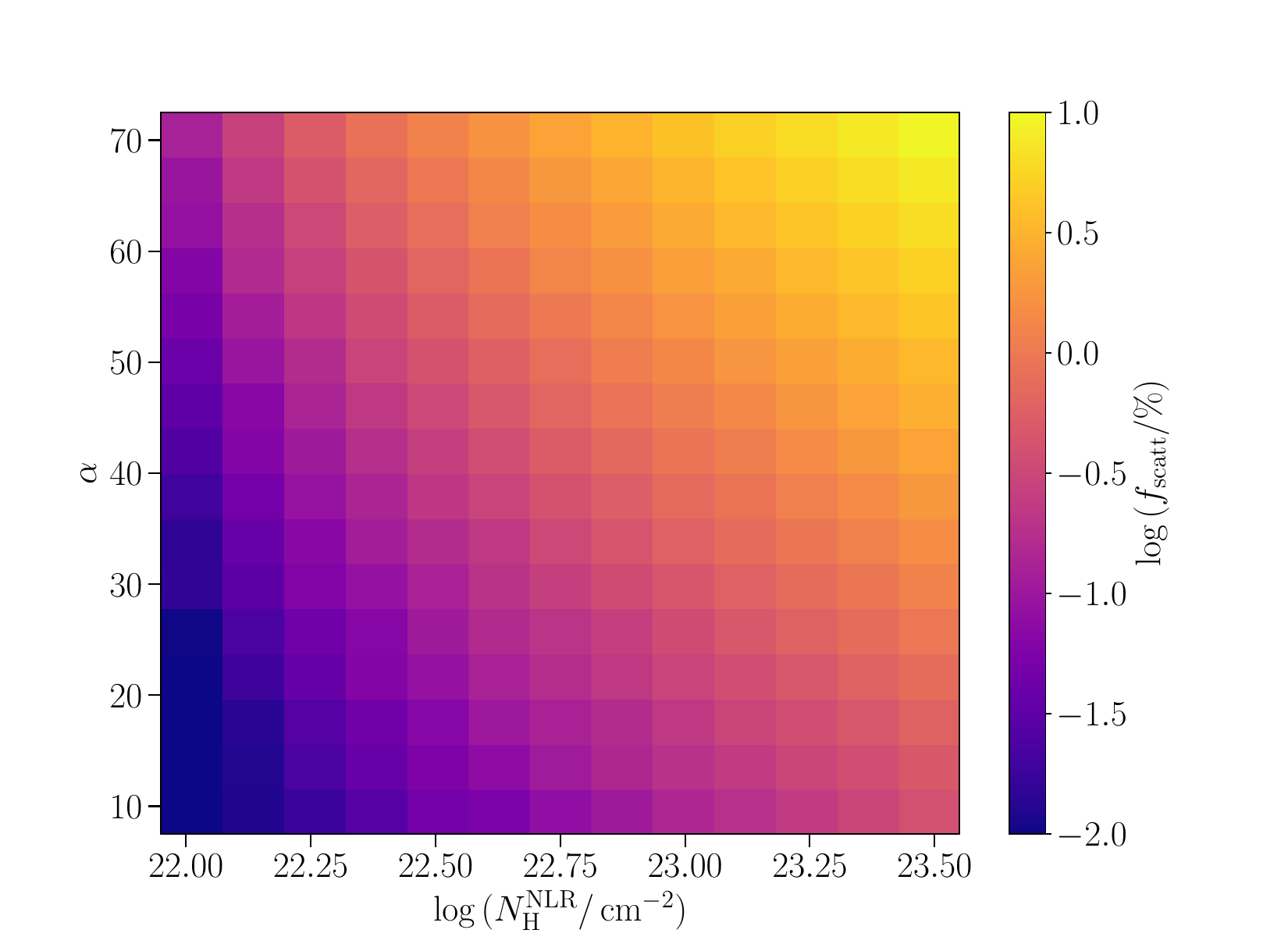}
                \caption{Contour plot showing the different opening angles ($\alpha = $10$-$70$^{\circ}$) and NLR column densities [$\log{(N_{\text{H}}^{\text{NLR}}\big/\text{\,cm}^{-2})} = 22.0-23.5$] explored by our \textsc{RefleX} simulations, and the corresponding $f_{\text{scatt}}$ value on the color bar. The data shown here are for a single inclination angle, $i=85^{\circ}$.}
                \label{fig3D}
            \end{figure}
            
            \subsection{Calculating the Scattering Fraction}\label{simulations}
            
                We set out to study the relationship between $f_{\text{scatt}}$ and the properties of a dusty torus surrounding the accreting system. These properties include the line-of-sight column density of the torus ($N_{\mathrm{H}}$) and its covering factor ($f_{\mathrm{cov}}$). G21 used the $f_{\text{scatt}}$ obtained by \cite{2017ApJS..233...17R}, where $f_{\mathrm{scatt}}$ is measured by fitting the spectrum of each obscured AGN with nine models in the spectral fitting package \textsc{XSPEC} including a cutoff powerlaw (\textsc{CUTOFFPL}), photoelectric absorption from neutral material (\textsc{ZPHABS}), a reflection component (\textsc{PEXRAV}) as well as a Thomson scattering component of the primary continuum. $f_{\mathrm{scatt}}$ was used as a free parameter to renormalise the flux.

                We calculate $f_{\mathrm{scatt}}$ using the following method. In a single simulation, \textsc{RefleX} can isolate the spectrum of photons which undergo only scattering interactions with a user defined resolution. In this study we choose 5\,eV, but this choice has little effect on our results. $f_{\mathrm{scatt}}$ is then calculated by dividing the Thomson scattered spectrum by the continuum flux in each energy bin and averaging the results. 
                
                In addition to the NLR column densities expected from observations found in $\S$\ref{density}, we also create a grid of \textsc{RefleX} calculations for a range of NLR column densities and half opening angles. These range from $\alpha$ = 10$-$70$^{\circ}$ in steps of $5^{\circ}$ and $\log{(N_{\text{H}}^{\text{NLR}}\big/\text{\,cm}^{-2})} = 22.0-23.5$ in steps of $0.1$ dex. This grid is created considering only a single inclination angle ($i$; also known as viewing angle, i.e., angle which the ``detector'' makes with the $z$ axis in Figure\,\ref{fig1}) of $85^{\circ}$. We calculate $f_{\mathrm{scatt}}$ for each choice of $N_{\mathrm{H}}^{\mathrm{NLR}}$ and $\alpha$. 
                
                When using the NLR column density obtained by averaging the \cite{2006A&A...459...55B} sources, we calculate $f_{\mathrm{scatt}}$ as a function of inclination angle ranging from $45^{\circ}-90^{\circ}$ in steps of $5^{\circ}$ where the method of calculating $f_{\mathrm{scatt}}$ above is repeated for each chosen inclination angle. We take care the detector only samples from inclinations which are larger than the half opening angle of the NLR to be consistent with the idea that we are simulating type\,2 (edge-on) AGN. This was done for the half opening angles $\alpha = 10, 30, \mathrm{and\,\,} 70^{\circ}$, in order to explore the effect of a varying NLR covering factor on the scattering faction.  Note, while we do not include a torus in our simulations, the half opening angle of the NLR determines the covering factor of the torus in the equatorial region. This is due to the fact that obscuring material is thought to collimate the NLR (e.g., \citealp{1985ApJ...297..621A, 1998ApJS..117...25M, 2013ApJS..209....1F, 2018ApJ...868...14S}). Thus, increasing values of $\alpha$ is equivalent to decreasing the covering factor of the torus [i.e., $f_\text{cov} = \cos{(\alpha)}$].

        \subsection{Theoretical Estimate of the Scattering Fraction}

            Here we derive a theoretical estimate of the amount of Thomson scattered radiation from an optically thin gas cloud to compare to our numerical results. The optical depth of photons to electron scattering can be calculated by $\tau_{e^{-}}= N_{\mathrm{H}}^{\mathrm{NLR}}\sigma_{\mathrm{T}}$, where $\sigma_{\mathrm{T}}\approx 0.665\times 10^{-24}\,$cm$^{2}$ is the Thomson cross-section. The amount of scattered radiation will be proportional to $\tau_{e^{-}}$, multiplied by the fraction of material available for scattering, i.e., $f_{\mathrm{cov}}$, as shown below:
            \begin{equation}\label{fscattau}
                f_{\mathrm{scatt}} \simeq N_{\mathrm{H}}^{\mathrm{NLR}}\sigma_{\mathrm{T}}(1 - f_{\mathrm{cov}}) = N_{\mathrm{H}}^{\mathrm{NLR}}\sigma_{\mathrm{T}}(1 - \cos{(\alpha)})  
            \end{equation}
            where $\alpha$ is the half opening angle of the NLR. We can see $f_{\mathrm{scatt}}$ is linearly dependant on both $N_{\mathrm{H}}^{\mathrm{NLR}}$ and $f_{\mathrm{cov}}$. Considering $\alpha = 45^{\circ}$ and $N_{\mathrm{H}}^{\mathrm{NLR}}=10^{22.43}\,\mathrm{cm}^{-2}$, we would expect a scattering fraction of $f_{\mathrm{scatt}} = 0.52\%$ from pure Thomson scattering. This is within the limits of what should be expected using the average $f_{\mathrm{scatt}}$ from \cite{2017ApJS..233...17R}.
    
    \section{Results}\label{results}

        \subsection{Numerical Results}\label{numr}

        \begin{figure*}
            \centering
            \includegraphics[scale=0.43]{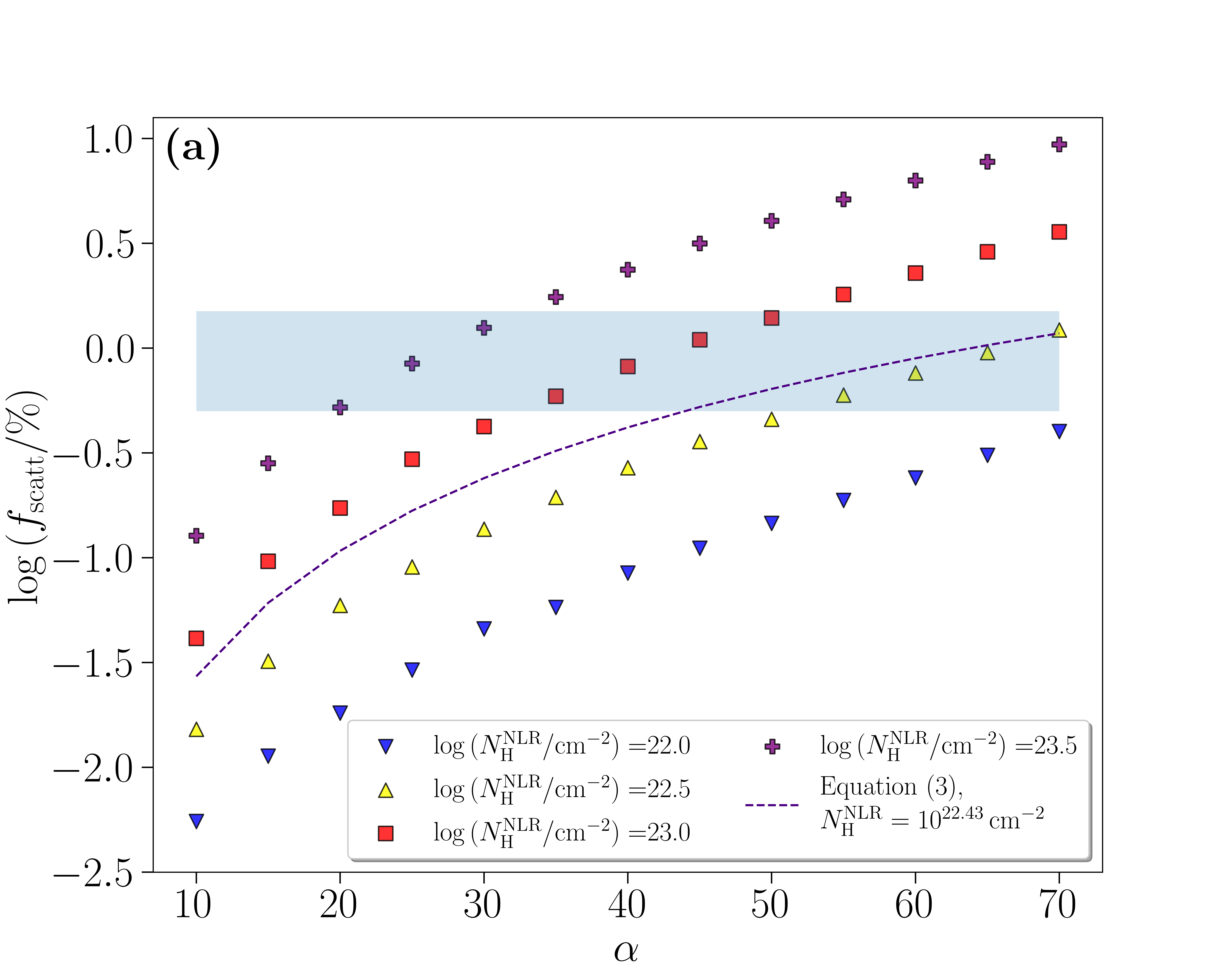}
            \includegraphics[scale=0.43]{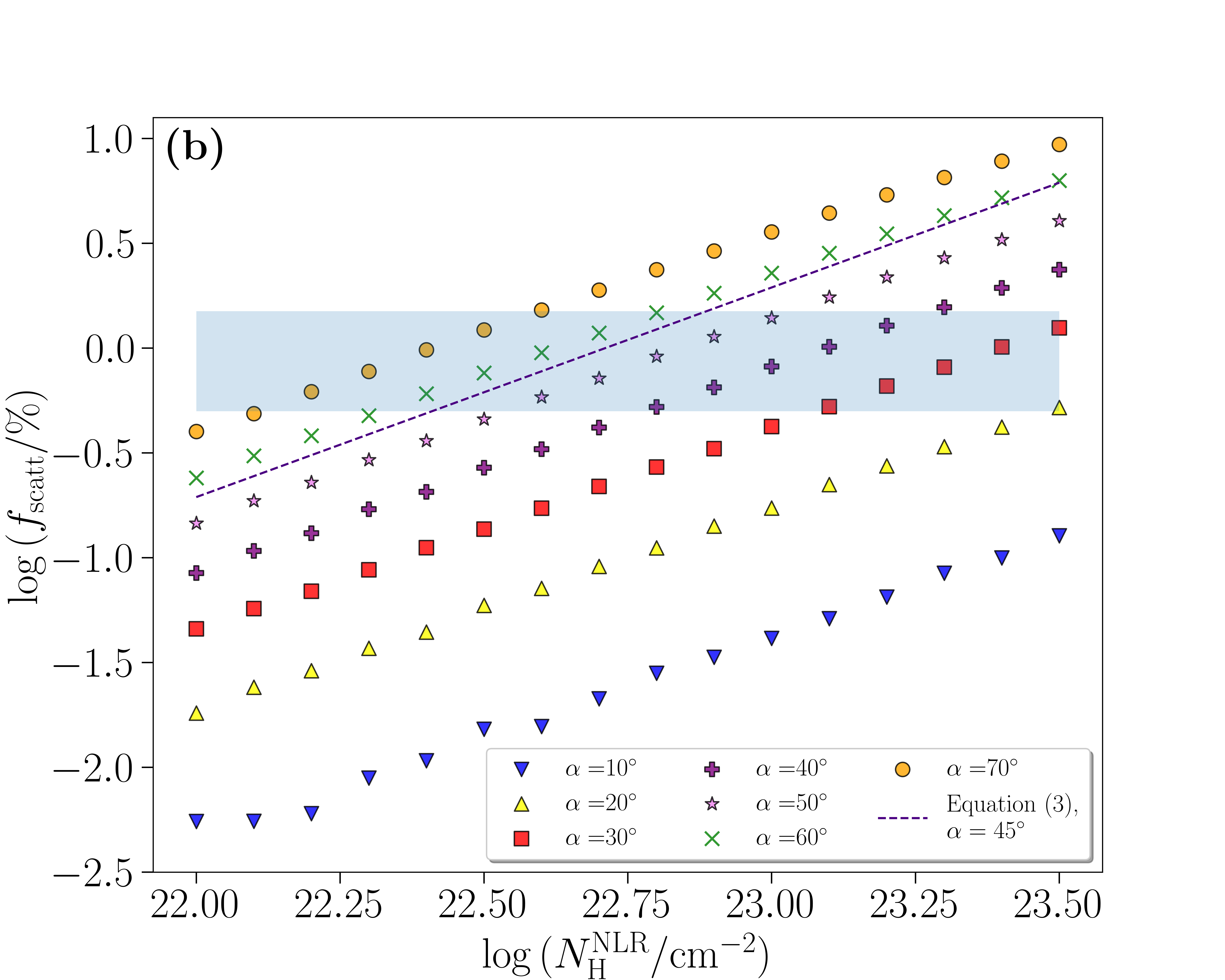}
            \caption{Panel \textbf{(a)}: scattering fraction for a NLR of varying column density as a function of the half opening angle at an inclination angle of $i=85^{\circ}$. Panel \textbf{(b)}: scattering fraction calculated for various NLR opening angles as a function of NLR column density. The blue shaded region represents the median value of the scattering fraction with errors from the {\it Swift}/BAT 70-month all-sky survey taken from \protect\cite{2017ApJS..233...17R}. The indigo dashed line shows the theoretical treatment in Equation\,\ref{fscattau}.}
            \label{fig3}
        \end{figure*}
    
        Using our \textsc{RefleX} simulations, we derive $f_{\text{scatt}}$ as a function of inclination angle for $\alpha = 10^{\circ}, 30^{\circ}$, and $70^{\circ}$. This was done for the averaged value of $N_{\mathrm{H}}^{\mathrm{NLR}}$ from \cite{2006A&A...459...55B} as discussed in $\S$\ref{simulations}. These results are illustrated in Figure\,\ref{fig2} while a contour plot of our full grid of simulations are shown in Figure\,\ref{fig3D}. Figure\,\ref{fig2} also includes the Monte Carlo (MC) linear regression line from G21 (black dashed line), as well as the MC simulation results per column density bin (purple points with grey error bars) from G21. It can be seen from Figure\,\ref{fig2} that $f_\mathrm{scatt}$ increases with $\alpha$ due to the increase in gas available for scattering. Some dependence on $i$ is expected due to the differential Thomson cross-section (e.g., G21), with photons being more likely scattered $0^{\circ}$ or $180^{\circ}$ to the incident radiation. However, our results show the correlation of $f_{\text{scatt}}$ with inclination angle is only very weakly negative, leading to significant discrepancies in the slopes of the simulated and observed relations. Therefore, based on these calculations, it is unlikely inclination angle alone accounts for the trend observed by G21.   
    
        Figure\,\ref{fig3D} shows a contour plot of $f_{\mathrm{scatt}}$ as a function of both $\alpha$ and $\log{(N_{\mathrm{H}}^{\mathrm{NLR}}\big/\mathrm{cm}^{-2})}$ for a inclination angle of $i=85^{\circ}$. This value of $i$ was chosen to be consistent with observations typical of the most inclined type 2 sources. For similar reasons as discussed above, $f_{\mathrm{scatt}}$ increases with both $\alpha$ and $\log{(N_{\mathrm{H}}^{\mathrm{NLR}}\big/\mathrm{cm}^{-2})}$ due to the increase in free electrons. This is better seen in Figure\,\ref{fig3}, which shows cross-sections of the contour plot. The shaded region is the median scattering fraction value ($f_{\text{scatt}} = 1.0\pm 0.5\% $) for the sample of AGN taken from the 70-month {\it Swift}/BAT all-sky survey \citep{2017ApJS..233...17R}. The theoretical estimate from Equation\,\ref{fscattau} is also shown in Figure\,\ref{fig3} as an indigo dashed line which agrees well with the trend and normalisation seen in both panels.   

        \begin{figure}
            \centering
            \includegraphics[scale = 0.33]{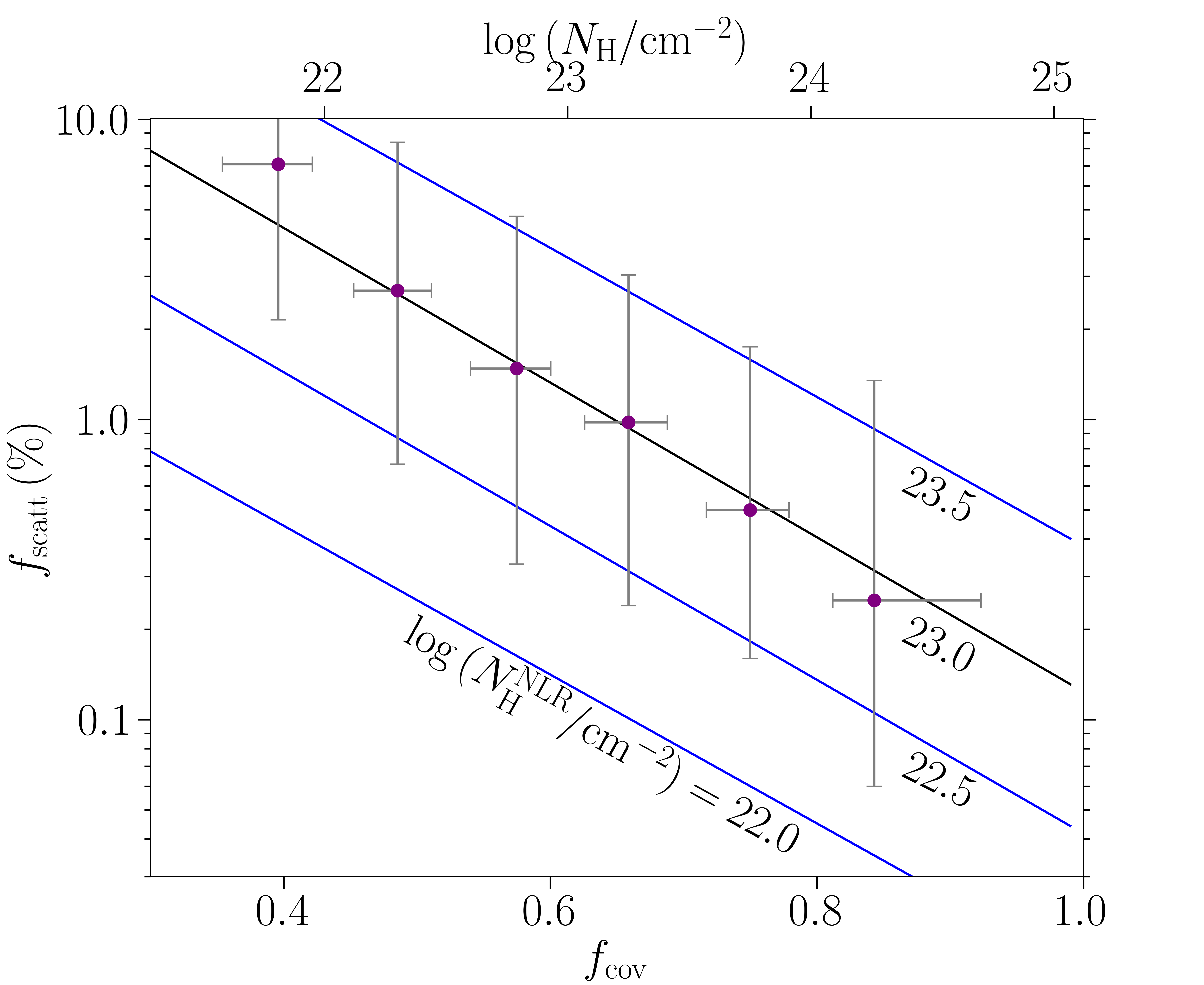}
            \caption{$f_\text{scatt}$ as a function of covering factor (bottom axis) and line-of-sight column density ($N_{\mathrm{H}}$; top axis) for chosen values of $N_{\mathrm{H}}^{\mathrm{NLR}}$ along with the mean values of $f_{\mathrm{scatt}}$ per $N_{\mathrm{H}}$ bin found in G21 (purple data points with error bars). The black trend line is the relation found by G21 reproduced by Equation\,\ref{eq4}. The Blue lines represent other possible choices of NLR column density which do not reproduce the G21 relation when using Equation\,\ref{eq4}.}
            \label{figfcov}
        \end{figure}

    \section{Discussion}\label{Discussion}
    
        \subsection{Normalisation and slope of the $f_{\mathrm{scatt}}-N_{\mathrm{H}}$ relation}
        
            \subsubsection{Normalisation}\label{norm}
        
            Here we discuss the values of $f_{\text{scatt}}$ obtained by our simulations as compared to the observed values reported in G21, and to the median value of a large sample of hard X-ray selected AGN from \cite{2017ApJS..233...17R}. As can be seen in Figure\,\ref{fig2}, the simulated scattered fraction only agrees with the observed values in  G21 for $\alpha\geq 30^{\circ}$, or $f_{\mathrm{cov}}\leq 0.87$. From our estimate shown in Equation\,\ref{fscattau}, we expect $f_{\mathrm{scatt}}$ to be dependant on both $f_{\mathrm{cov}}$ (and thus $\alpha$) and $N_{\mathrm{H}}^{\mathrm{NLR}}$. This can be understood physically due to the fact that as the opening angle increases, keeping the column density fixed causes an increase in the number of free electrons available for scattering. This is shown in our numerical calculations in panel \textbf{(a)} of Figure\,\ref{fig3} where, for a chosen inclination angle of $i=85^{\circ}$ and fixed NLR column density, $f_{\text{scatt}}$ increases as a function of the opening angle $\alpha$. Additionally, higher column densities would also increase the number of available electrons. We investigate this in panel \textbf{(b)} of Figure\,\ref{fig3}, where the column density of the NLR is allowed to vary. This shows that NLR column densities of $\log{(N_{\text{H}}^{\text{NLR}}\big/\text{\,cm}^{-2})}\geq 22.2$ are needed to achieve the required normalisation from \cite{2017ApJS..233...17R}, depending on the opening angle. Varying the height of the NLR would likely have a similar effect. Equation\,\ref{fscattau}, shown as the dash indigo line, broadly agrees with our numerical results.
            
            What could lead to different sources having different values of $N_{\mathrm{H}}^{\mathrm{NLR}}$? There could be differences in the free electron column densities of different AGN potentially associated to different SEDs. An SED with a higher Eddington ratio might lead to a higher UV flux relative to the X-ray flux, and thus create more free electrons (e.g., \citealp{2012MNRAS.420.1848D}). However, \cite{2010ApJ...711..144N} and G21 found a slight negative dependence of $f_\mathrm{scatt}$ on $\log{(L/L_{\text{Edd}})}$, which might suggest that in rapidly accreting AGN the radiative force has expelled all gas responsible for the soft X-ray excess. 
            
            It should be noted that we have also neglected additional physical processes which could significantly contribute to the fraction of soft X-ray radiation such as photoionized emission lines. Photoionized lines dominate the soft X-ray emission in many observations of type\,2 AGN with non-detectable continuum in the $\sim$0.1$-$2\,keV range (e.g., \citealp{2003ApJ...596..114S, 2006A&A...448..499B, 2007MNRAS.374.1290G}). Radiative recombination continuum features (e.g., \citealp{2006A&A...448..499B,2007MNRAS.374.1290G}) could also contribute extra scattered flux and increase the normalisation of $f_{\text{scatt}}$ relative to what is expected when considering Thomson scattering alone. Additionally, free electrons from the powerful ionised outflows produced in the accretion disk (e.g., \citealp{Tombesi:2010oj,Fukumura:2010og,Harrison:2018fh}) could contribute to the column density of the scattering material in the polar region of the system, therefore enhancing the fraction of scattered photons. However, our results only attempt to explain the relation found in G21 in the context of Thomson scattering as proposed by G21 with no contribution from these processes. If, for example, photoionized lines were included in our calculations, they are likely to contribute a proportional factor due to the possibility that emission lines and Thomson scattered radiation could be co-spatial. 
            
            Fluorescent line emission is also expected in the X-ray spectrum of AGN as seen most notably in the Fe\,K$\alpha$ line at $\sim$6.4\,keV (e.g., \citealp{2000PASP..112.1145F, 2004ApJ...604...63Y, 2010ApJS..187..581S, 2011ApJ...727...19F, 2014MNRAS.441.3622R}). It has been shown that the new polar component recently discovered in the Mid-Infrared emission of many local AGN (e.g., \citealp{2012ApJ...755..149H, 2013A&A...558A.149B, Asmus_2016, refId03, Leftley_2019}) can lead to the production of fluorescent lines in the $\sim$0.3$-$5\,keV range \citep{2022MNRAS.512.2961M}. Thus this component could possibly increase the normalisation of the scattered emission. To estimate the increase in $f_{\mathrm{scatt}}$ resulting from this component, we integrate the total contribution from all the lines found in \cite{2022MNRAS.512.2961M} and find, at inclination angles of $80^{\circ}-90^{\circ}$, this component could contribute up to 0.126$\%$ to the scattered flux. However, this contribution is subject to change based on the column density, and extent of the polar region (see also \citealp{Liu2019XraySO}).   
            
            \subsubsection{Reproducing the $f_{\mathrm{scatt}}-N_{\mathrm{H}}$ slope from G21}\label{slope}
            
            The dashed black line in Figure\,\ref{fig2} shows the relationship between $f_{\text{scatt}}$ and the line-of-sight column density observed in G21 (top axis). In order to explain this correlation, G21 drew a relation between the torus column density and the inclination angle of a source. This was done by assuming that the circumnuclear material in all sources of their sample had a similar geometry. This geometry is a SMBH surrounded by an equatorial torus with different vertical layers having different column densities (increasing with inclination angle; see Figure\,6 in G21). This implies the only parameter that changes between sources is the inclination angle. The covering factors of different layers of the torus were obtained by following the statistical arguments given in \citet{2015ApJ...815L..13R} who studied a large sample of AGN and the relationship between line of sight $N_{\mathrm{H}}$ and covering factor. Ricci et al. took advantage of the fact that the probability of observing an obscured AGN with $\log{(N_{\mathrm{H}}\big/\mathrm{cm}^{-2})}\geq22$ is directly related to its covering factor. Thus, the measure of a distribution of covering factors and line-of-sight column densities will provide a relationship between $N_{\mathrm{H}}$ and $i$ when assuming $i = \arccos{(f_{\mathrm{cov}})}$ (G21). It was also assumed that the denser material resides closer to the equatorial plane of the torus. \citet{2015ApJ...815L..13R} found 70$\%$ of their sample was obscured, meaning that the average covering factor of gas with $\log{(N_{\mathrm{H}}\big/\mathrm{cm}^{-2})}\geq22$ in the AGNs in their sample was 70$\%$. This will correspond to an inclination angle of $i\gtrsim 46^{\circ}$ for $\log{(N_{\mathrm{H}}\big/\mathrm{cm}^{-2})}\geq22$. Additionally, covering factors of $52\%$ and $27\%$ were found on average for column densities $\log{(N_{\text{H}}}\big/\text{\,cm}^{-2}) \geq 23$ and $\log{(N_{\text{H}}}\big/\text{\,cm}^{-2}) \geq 24$, respectively. These column densities correspond to inclination angles of $i \gtrsim 59^{\circ}$ and  $i \gtrsim 74^{\circ}$, respectively. We use these points to construct a linear relation between the inclination angle (in degrees) and the torus column density. This relationship is shown below in Equation\,\ref{eq1p5}
            \begin{equation}\label{eq1p5}
                \log{(N_{\text{H}}\big/\text{\,cm}^{-2})} = 19.24 + 0.07i.
            \end{equation}
            Using this equation, we change variables to plot $f_{\text{scatt}}$ as a function of $\log{(N_{\text{H}}\big/\mathrm{cm}^{-2})}$ on the top x-axis of Figure \ref{fig2}. Comparing the results of our simulations with the observational results of G21 (dashed black line in Figure\,\ref{fig2}) it is obvious that the dependence of $f_{\rm scatt}$ on the inclination angle based on Equation\,\ref{eq1p5} cannot reproduce the slope of the observed negative correlation between $f_{\rm scatt}$ and $N_{\rm H}$ and another explanation is needed.
            
            A second possible explanation for the slope of the $f_{\rm scatt}-N_{\rm H}$ relation is that the average covering factor of the circumnuclear material in sources with higher column densities tend to be larger than in less obscured sources. This would be in agreement with what was previously proposed by \cite{2007ApJ...664L..79U} to explain the existence of sources with extremely low values of $f_{\rm scatt}$. Moreover, several observational studies both in the IR and X-rays have suggested that obscured type\,2 sources with high column densities also have higher covering factors (e.g., \citealp{2011ApJ...731...92R, 2011A&A...532A.102R, 2016ApJ...819..166M, 2018ApJ...853..146T, 2019A&A...626A..40P}; G21). Here we use the results of our simulations to assess what relation between the covering factor and the torus column density would be needed to explain the $f_{\rm scatt}-N_{\rm H}$ relation. 

            Figure\,\ref{fig3D} shows that in addition to the column of free electrons, the scattering fraction also depends on the opening angle of the NLR (i.e., $f_{\text{cov}}$; also see Equation\,\ref{fscattau} and panel \textbf{(a)} of Figure\,\ref{fig3}). In order to find a relation between the torus column density and its covering factor, we first consider a NLR at an inclination of $85^{\circ}$ to the observer and column density of $\log{(N_{\text{H}}^{\text{NLR}}\big/\text{\,cm}^{-2})} = 23.0$ at various opening angles $\alpha$. We then assume a linear relation between $f_{\mathrm{scatt}}$ and $f_{\mathrm{cov}}$ by using the fact that $f_{\mathrm{cov}} = \cos{({\alpha})}$. This gives  
            \begin{equation}\label{eq3p5}
                \log{(f_{\mathrm{scatt}}\big/\%)}=1.67 - 2.57f_{\mathrm{cov}}.
            \end{equation}
            We assume the relationship between $f_{\mathrm{scatt}}$ and $f_{\mathrm{cov}}$ from our calculations should give the same results as found in G21 and thus Equation\,\ref{eq3p5} was set equal to the $f_{\mathrm{scatt}}-N_{\mathrm{H}}$ trend found by G21. Solving for $f_{\mathrm{cov}}$ yields
            \begin{equation}\label{eq4}
                f_{\text{cov}} = -3.59 + 0.18\log{(N_{\text{H}}\big/\text{\,cm}^{-2})}.
            \end{equation}
            Using this relationship to translate between $f_{\text{cov}}$ and $N_{\text{H}}$, we plot in Figure\,\ref{figfcov} the scattering fraction as a function $f_{\text{cov}}$ and $N_{\text{H}}$ (bottom and top axes, respectively), along with the median value data points of $f_{\text{scatt}}$ per column density bin from G21. Using Equation\,\ref{eq4}, we are able to perfectly reproduce the best fit line to the data found in G21. This is shown as the black line in Figure\,\ref{figfcov}. Other values of $N_{\mathrm{H}}^{\mathrm{NLR}}$ which fail to reproduce the G21 relation using Equation\,\ref{eq4} are shown in blue. Thus, a positive correlation between $f_{\text{cov}}$ and $N_{\text{H}}$ could easily explain the $f_{\mathrm{scatt}}-N_{\mathrm{H}}$ trend. Now, when using this relation to translate covering factors into observed column densities, one must note that, because covering factors cannot be negative or exceed unity, the valid range of column densities that Equation\,\ref{eq4} can be used is $\log{(N_{\text{H}}\big/\text{\,cm}^{-2})} = [19.89, 25.44]$.    
        
        \subsection{Co-spatial partially and fully ionised regions}\label{region}

            Since we set out to study the soft X-ray excess that originates from  scattered radiation off free electrons in the NLR (e.g., \citealp{1997ApJ...488..164T, 2006A&A...446..459C,2007ApJ...664L..79U,2017ApJS..233...17R}) our simulations consists only of fully ionised Hydrogen and Helium providing the population of electrons responsible for the scattered X-ray radiation. However, photoelectric absorption and transitions from bound electrons of metals such as Oxygen and Silicon can absorb some the the X-ray emission (e.g., \citealp{2006agna.book.....O}). To test the impact of these bound electrons in our simulations, we consider the X-ray ionisation parameter given by:
            \begin{equation}
                \xi = \frac{4\pi F_{\mathrm{ion}}}{n_{\mathrm{H}}}=\frac{L_{\mathrm{ion}}}{n_{\mathrm{H}} r_{0}^{2}}
            \end{equation}
            where $n_{\mathrm{H}}$ is the Hydrogen number density, $r_{0}$ is the distance from the X-ray source, and $L_{\mathrm{ion}}$ is the ionising luminosity. Estimating {\bf $L_{\mathrm{ion}}= 10^{44}\,\mathrm{erg\,s}^{-1}$}, $r_{0}= 1\,$pc and considering a range of densities of  $n_{\mathrm{H}}=10-10^{4}\,$cm$^{-3}$ gives a possible X-ray ionisation parameter of {\bf $\xi=10^{3}-10^{6}\,\mathrm{erg\,cm\,s}^{-1}$}. Thus, we expect that, close to the SMBH, most metals would be ionised and would not absorb significantly the X-ray radiation, particularly below 2\,keV. 
            
\begin{figure}
            \centering
            \includegraphics[scale = 0.3]{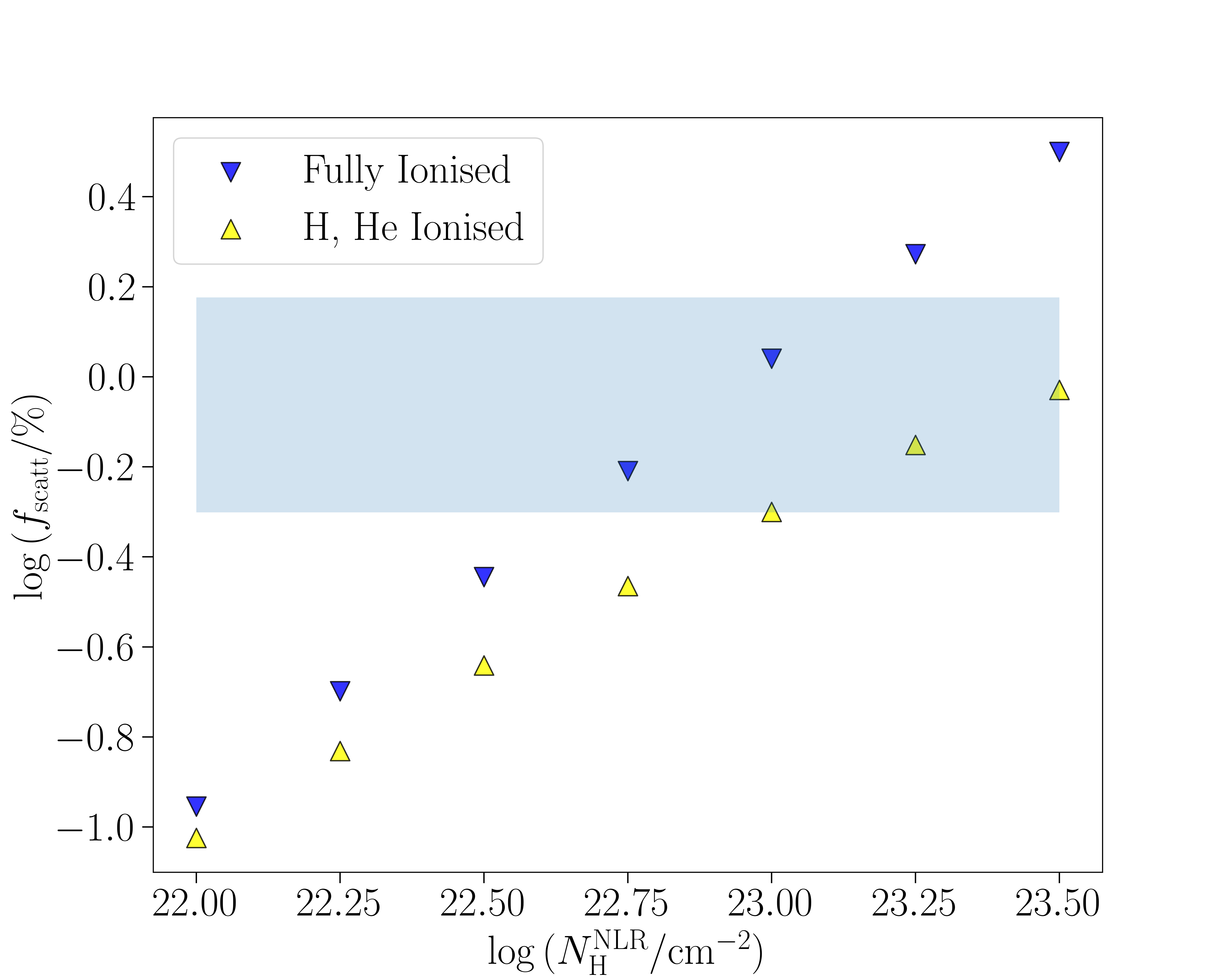}
            \caption{$f_{\mathrm{scatt}}$ versus $\log{(N_{\mathrm{H}}^{\mathrm{NLR}}\big/\mathrm{cm}^{-2})}$ for solar metallicity material with Hydrogen and Helium ionised, and fully ionised material. The partially ionised material could produce [O\,{\sc iii}] $\lambda$5007 emission while the fully ionised material gives rise to the scattered emission. The shaded region represents the median scattering fraction with errors from \protect\cite{2017ApJS..233...17R}.}
            \label{M1T0}
        \end{figure}

            However, there must also exist partially ionised metals in the NLR due to the fact that extended X-ray emission can be traced by [O\,{\sc iii}] $\lambda$5007 emission (e.g., \citealp{2001ApJ...556....6Y, 2006A&A...448..499B, 2010A&A...516A...9D}; G21). Indeed, the correlation between $f_{\mathrm{scatt}}$ and the $L_{X}/L$([O\,{\sc iii}] $\lambda$5007) ratio found by \cite{2015ApJ...815....1U} and G21 also confirms the idea that these regions are co-spatial. Partially ionised material would also absorb some of the X-ray emission through photoelectric processes. To study how the presence of partially ionised material changes the amount of scattered radiation, we plot in Figure\,\ref{M1T0} the scattered fraction as a function of column density from two NLRs, one consisting only of free electrons and the other consisting of solar metallicity material (abundances taken from \citealp{2003ApJ...591.1220L}) with only hydrogen and helium fully ionized. In \textsc{RefleX}, this is accomplished with various combinations of the \textsc{metallicity} and \textsc{temperature} commands. The \textsc{metallicity} command determines the composition of the material, \textsc{metallicity} 0 for hydrogen and helium only and \textsc{metallicity} 1 for solar abundances. The \textsc{temperature} command determines the ionisation state of the gas, \textsc{temperature} 0 for cold neutral material and \textsc{temperature} 1 for hot material where hydrogen and helium are fully ionised. Thus, material consisting of free electrons is achieved with `\textsc{metallicity} 0 \textsc{temperature} 1' and solar metallicity material with hydrogen and helium ionised is given by `\textsc{metallicity} 1 \textsc{temperature} 1'. 
            
            It can be seen in Figure\,\ref{fig3} that $f_\mathrm{scatt}$ from the fully ionised material is always a factor of $\sim$2 larger, and higher column densities of partially ionised of material are needed to reproduce the average $f_{\mathrm{scatt}}$ as seen from \cite{2017ApJS..233...17R} (shaded region). This, along with the expected presence of a photoelectric cutoff when bound electrons are present, implies that highly ionised material would produce significantly more scattered radiation in the soft X-rays. If the X-ray emission can traced with the [O\,{\sc iii}] $\lambda$5007, why does partially ionised material not account for the observed $f_{\mathrm{scatt}}$?   
            
            When clouds of gas in the NLR are directly exposed to the AGN SED, absorption of ionising photons will lead to compression of the photoionised layer (e.g., \citealp{2002ApJ...572..753D, 2004ApJS..153....9G, 2004ApJS..153...75G, 2014MNRAS.438..901S}). This Radiation Pressure Confinement (RPC; \citealp{2014MNRAS.438..901S}) leads to a highly ionised surface which emits X-rays, and a layer closer to the Hydrogen ionisation front which emits optical lines such as [O\,\textsc{iii}] $\lambda$5007. RPC also results in a thin photoionised region, many orders of magnitude smaller than its distance to the photoionising source \citep{2014MNRAS.438..901S}. Thus, RPC predicts a common origin for optical and soft X-ray emitting regions (\citealp{2014MNRAS.438..901S}). \cite{2019MNRAS.485..416B} studied the effects of radiation pressure confinement on soft X-ray emitting gas in a sample of obscured AGN using the \textit{XMM-Newton} Reflection Grating Spectrometer. They found that radiation pressure compression in the photoionised region reproduces the observed differential emission measure slope, implying the NLR in these AGN are indeed dominated by radiation pressure as opposed to another confinement mechanism such as a constant gas pressure multiphase medium. It is therefore possible that the illuminated faces of radiation pressure confined slabs, where the estimates made from Equation\,\ref{eq4} are true, are mostly likely what is driving the scattered soft X-ray emission observed in G21. These slabs inevitably reach the Hydrogen ionisation front where $\xi \ll 1$, the material becomes neutral and is no longer able to efficiently scatter the X-ray radiation below $\sim$2\,keV. This will explain the spatial extent and overlap between the X-ray and optically emitting regions. This highly ionised surface could also be the origin of spectra from species of iron such as Fe\,\textsc{xxv} and Fe\,\textsc{xxvi} which have been observed in Compton-thick AGN (e.g., \citealp{2002A&A...387...76B, 2005MNRAS.357..599B}).
            
    \section{Summary and conclusions}\label{conclusions}

        We have used X-ray simulations on the platform \textsc{RefleX} to study the origin of the soft X-ray excess in obscured AGN. G21 observed a negative correlation between the fraction of scattered radiation relative to the primary continuum with a sources column density. We attempt to discern between two possible explanations for this observation: (1) the angle dependence on the differential Thompson cross-section and (2) objects with higher column densities have covering factors closer to unity, as suggested by previous studies (e.g., \citealp{2007ApJ...664L..79U}; G21). We consider a full grid of opening angles and free electron column densities from the NLR, where most of the X-ray scattering takes place as traced by the [O\,{\sc iii}] $\lambda 5007$ emission (e.g., \citealp{2006A&A...448..499B}, G21). We also consider values derived using observations from \cite{2006A&A...459...55B} as well as {\sc Cloudy} calculations. While we assumed here that Thomson scattering in the NLR is the dominating process in the soft X-rays,  there could be significant contributions to the scattered component from photoionised lines or recombination continuum which we do not consider here. However, such components would be expected to be cospatial with the gas responsible for the Thomson scattered radiation. Figure\,\ref{fig2} and \ref{fig3D} show the main results of our simulations. Our main conclusions are the following:
        \begin{itemize}
            \item  We find a very weak negative correlation between $f_{\text{scatt}}$ and the inclination angle (see Figure \ref{fig2}). When using statistical arguments from \cite{2015ApJ...815L..13R} to translate between inclination angles and observed column densities, we are unable to produce the observed slope from G21. This, combined with the weak dependence of $f_{\text{scatt}}$ on the inclination angle, implies that a change in inclination angles alone cannot reproduce the relation found in G21. 
            \item The column densities expected for the NLR from observations only reproduce the observed values of $f_{\rm scatt}$ for a small range of $\alpha$ and $i$. We explored possible scenarios in which extra scattering can contribute to $f_{\mathrm{scatt}}$. This extra scattered flux could come from a higher column density of the NLR (see Figure\,\ref{fig3}), from additional radiative recombination continuum features (e.g., \citealp{2006A&A...448..499B,2007MNRAS.374.1290G}), extra free electrons from highly ionised material associated to outflows from the accretion disk (e.g., \citealp{Tombesi:2010oj,Fukumura:2010og,Harrison:2018fh}), photoionized lines seen in the soft X-rays of obscured type\,2 AGN using high spectral resolution instruments (e.g., \citealp{2003ApJ...596..114S, 2006A&A...448..499B, 2007MNRAS.374.1290G}), as well as from the polar component seen in the MIR of many local AGN, (e.g., \citealp{2012ApJ...755..149H, 2013A&A...558A.149B, Asmus_2016, refId03, Leftley_2019}) which has been shown to contribute many fluorescent lines to the soft X-ray spectrum of AGN (\citealp{Liu2019XraySO, 2022MNRAS.512.2961M}). This latter component can produce up to 0.126$\%$ of the primary continuum.
            \item  Using our full grid of simulations shown in Figure\,\ref{fig3D}, we showed that a positive relationship between line of sight column densities and covering factors would be able to reproduce the relationship between $f_{\text{scatt}}$ and $N_{\text{H}}$ observed in G21. The relationship can be seen in Equation\,\ref{eq4} and is plotted as the black line in Figure\,\ref{figfcov}. This implies that the main cause of the observed slope found in G21 might be that the covering factor in local AGN is correlated to its column density, with more obscured AGN also having larger covering factors of the circumnuclear material.   
        \end{itemize}

\section*{Acknowledgements}
The authors are very grateful to the anonymous referee for their insightful comments and suggestions on the paper. JM acknowlages support from NASA grant 80NSSC20K0038. CR acknowledges support from the Fondecyt Regular grant 1230345 and ANID BASAL project FB210003. 

\section*{Data Availability}
The datasets generated and/or analysed in this study are available from the corresponding author
on reasonable request.




\bibliographystyle{mnras}
\bibliography{bibi} 




\appendix

\section{Importance of the Electron Density Distribution}\label{AA}

\begin{figure}
        \centering
        \includegraphics[scale = 0.29]{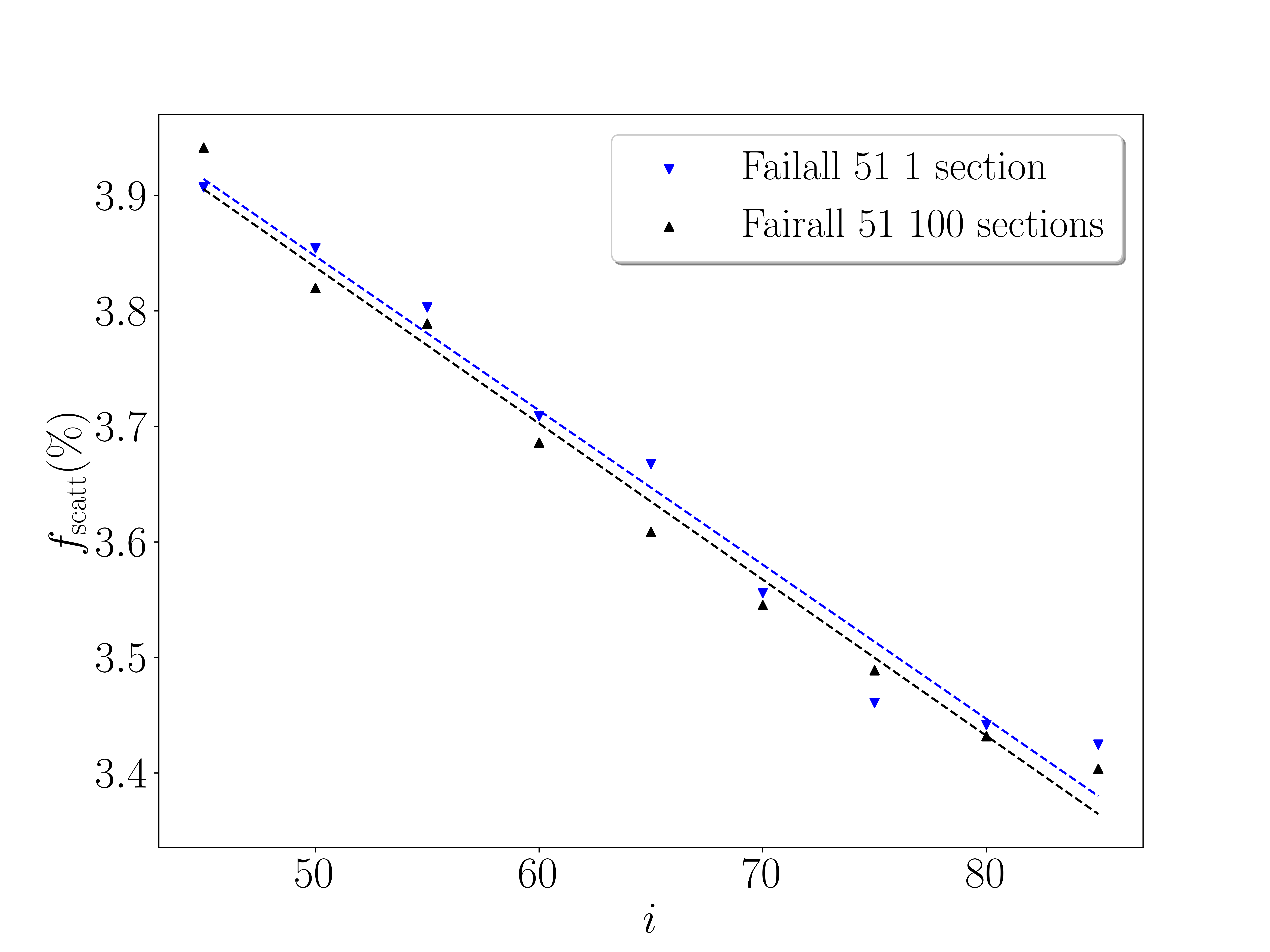}
        \includegraphics[scale = 0.29]{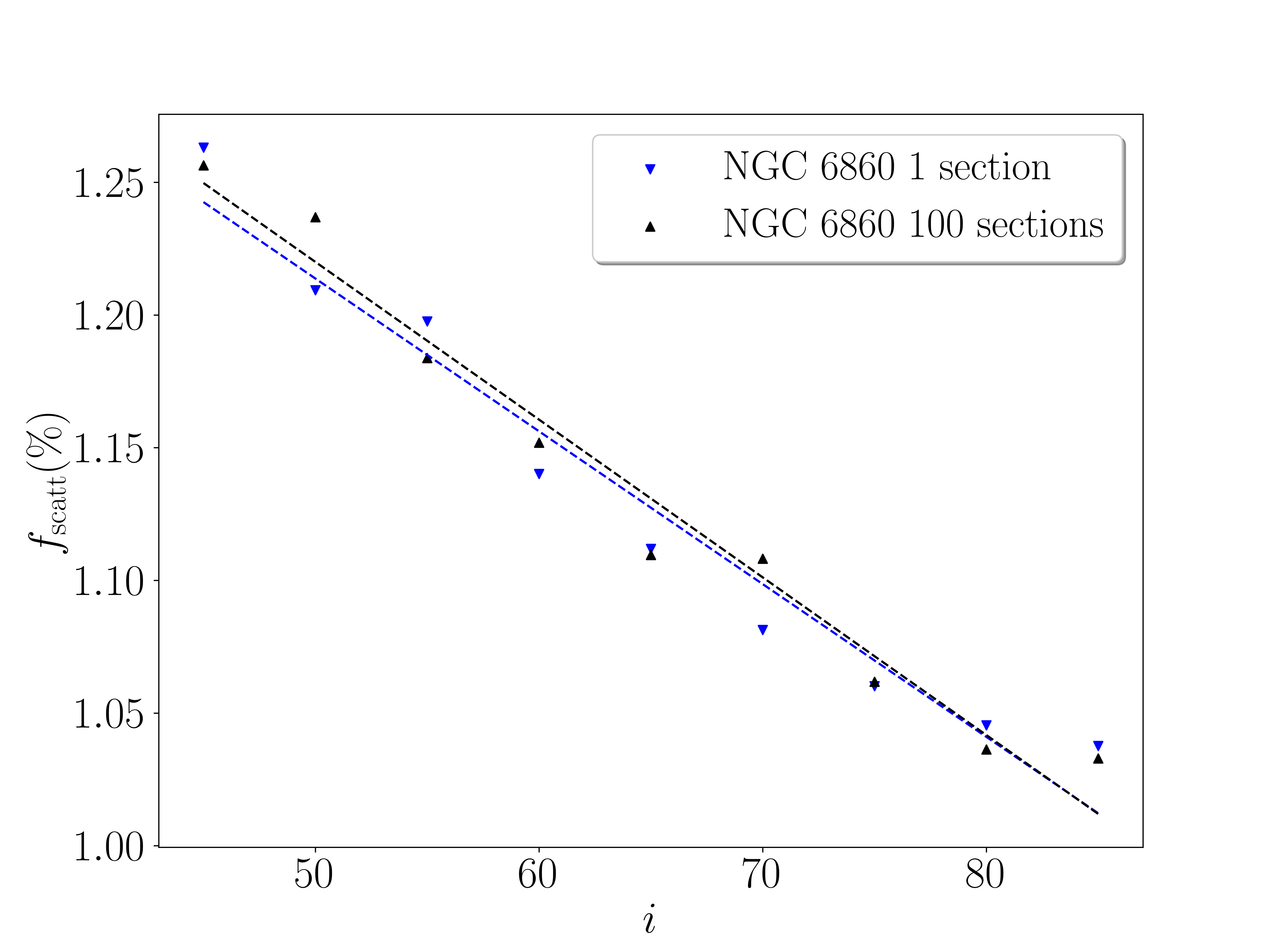}
        \includegraphics[scale = 0.29]{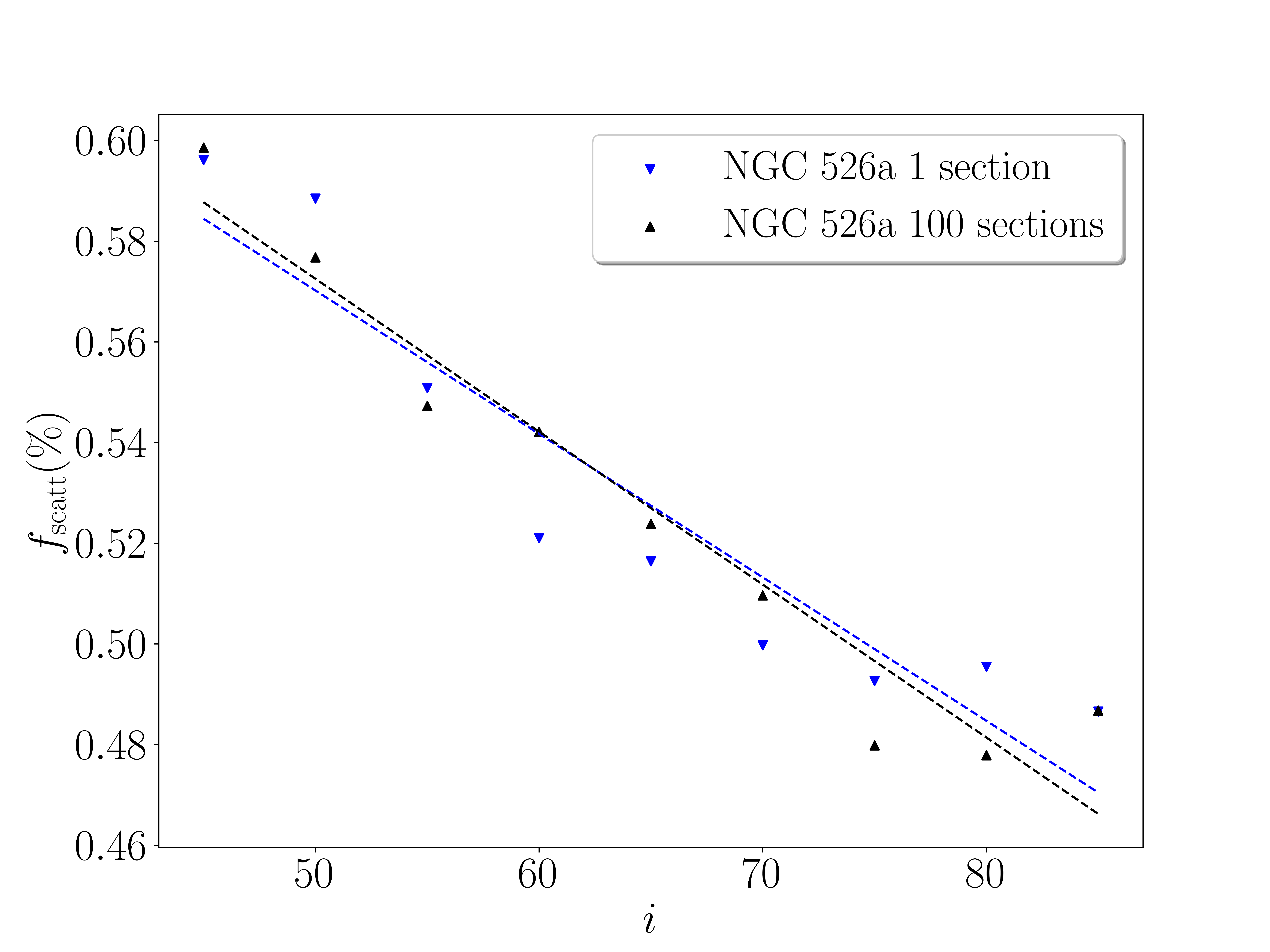}
        \caption{Scattering fraction as a function of inclination angle for three of the AGN from \protect\cite{2006A&A...459...55B}, calculated by using either the full prescription of Equation\,\protect\ref{eq1} (black upside-down triangles; best fit trend shown as black dashed line) or a single, constant density NLR with the same overall column density (black triangles; best fit trend shown as blue dashed line). Each panel in the figure represents the scattering fraction as a function of the inclination angle for Fairall 51 (upper panel), NGC 6860 (middle panel), and NGC 526a (lower panel).}
        \label{figA1}
    \end{figure}

\begin{figure}
        \centering
        \includegraphics[scale = 0.29]{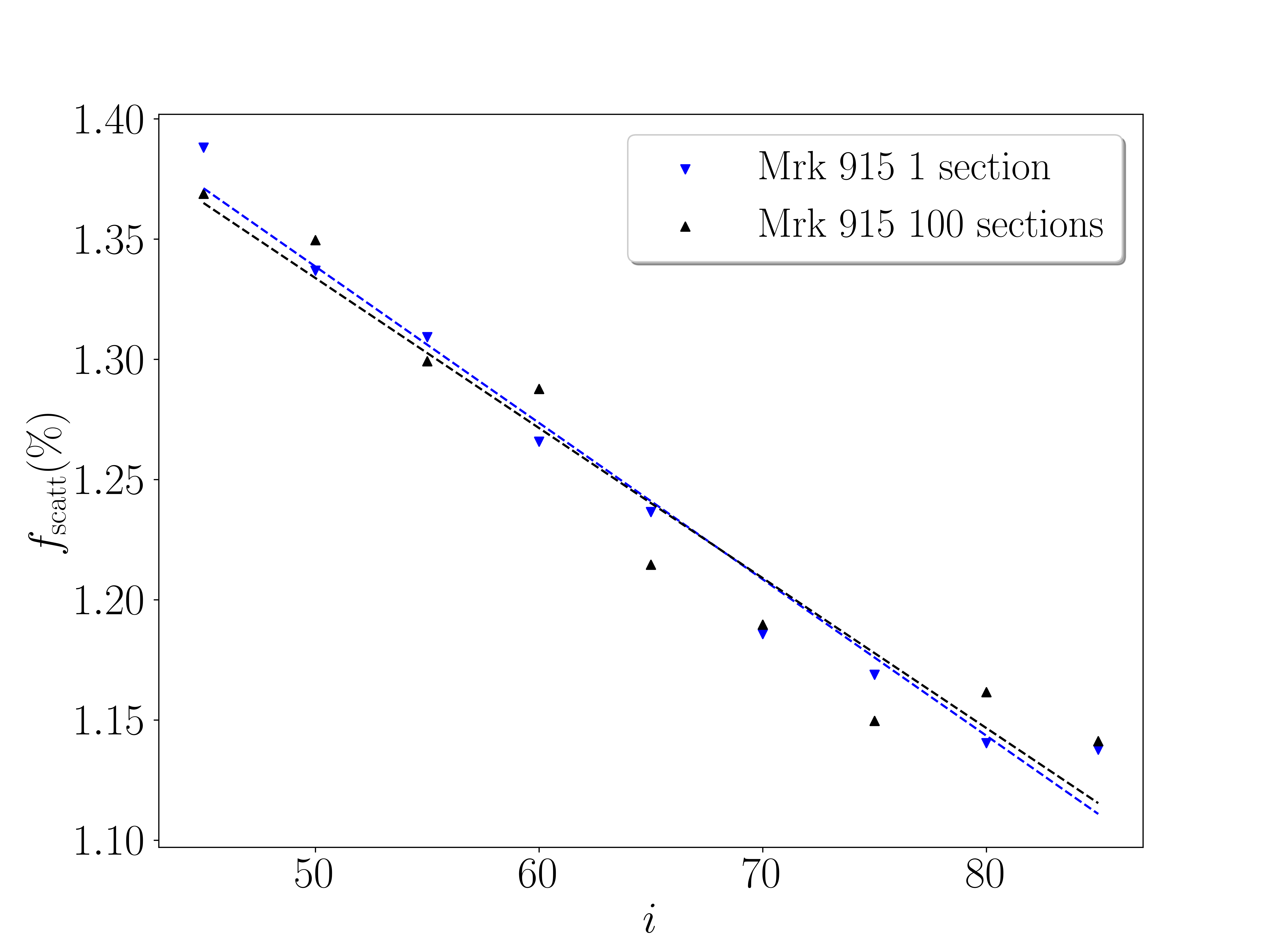}
        \includegraphics[scale = 0.29]{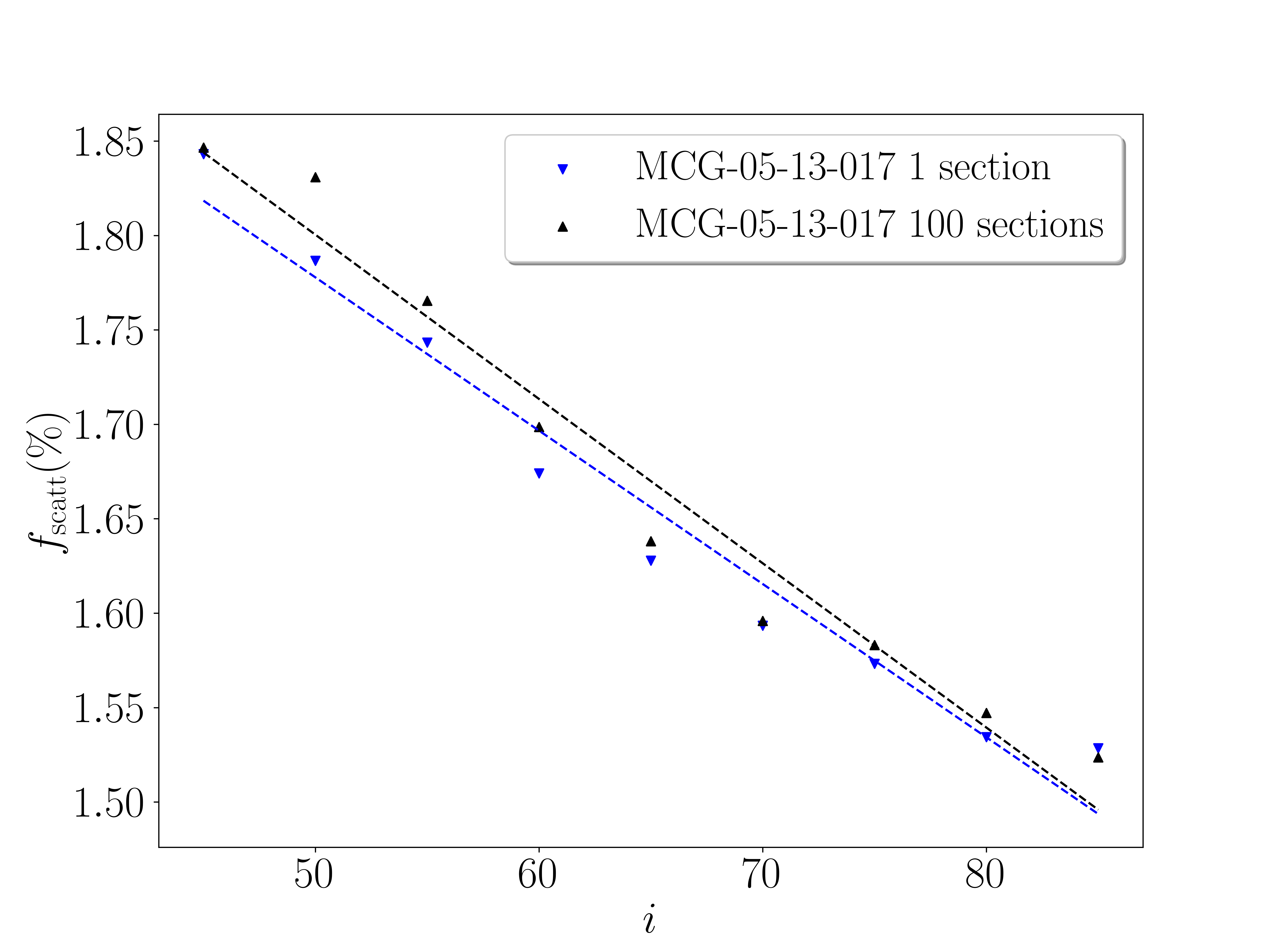}
        \includegraphics[scale = 0.29]{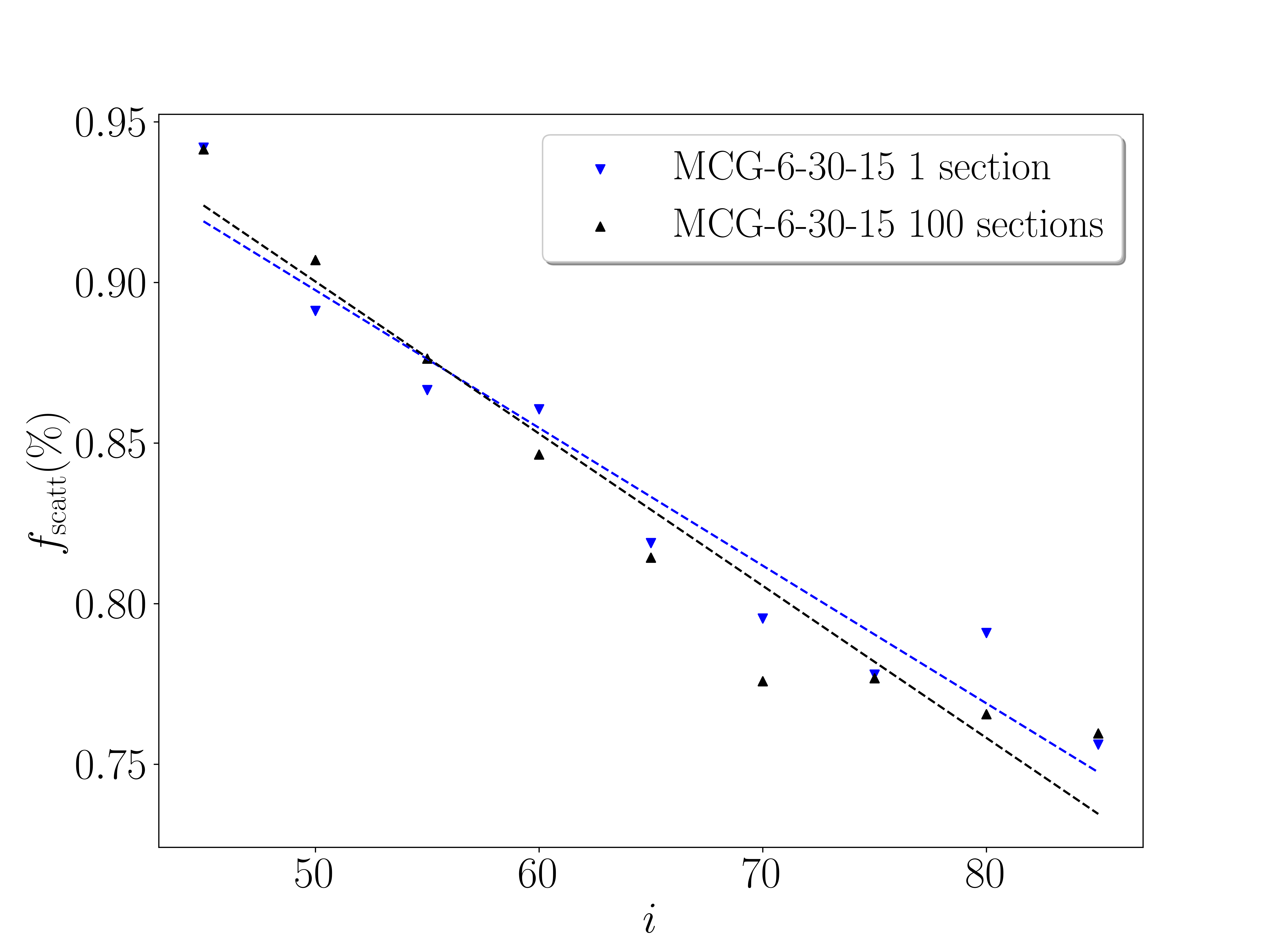}
        \caption{Same as Figure\,\ref{figA1} but for Mrk 915 (upper panel), MCG -05-13-017 (middle panel), and MCG -06-30-15 (lower panel)}
        \label{figA2}
    \end{figure}

    In $\S$\ref{density} we state that the exact distribution of free electrons with distance is unimportant when calculating the scattering fraction and the only parameter which needs to be considered is the total column density of the NLR. We show this in Figures\,\ref{figA1} and \ref{figA2}. In these figures, we use the six sources from \cite{2006A&A...459...55B} and implement the electron density distribution in Equation\,\ref{eq1} by splitting up the NLR into 100 constant density sections and calculating the density at the centre of the section. The scattering fraction as a function of inclination angle, when considering this exact density distribution, can be seen as the black triangles in Figures\,\ref{figA1} and \ref{figA2} with the best fit curve being the dashed black line for each source. The blue triangles in the figure represent the scattering fraction at different inclinations from a NLR consisting of only one constant density section but with the same overall column density as its counterpart with 100 sections. Each data point as well as the best fit curve shows that, for each source, the scattering fraction is similar irrespective of using the exact electron distribution seen in Equation\,\ref{eq1}, or the average column density for that particular source.         

    
\section{Parameters for {\sc Cloudy} Simulations}\label{AB}

Here we describe our {\sc Cloudy} simulations to infer another value for the free electron column density of the NLR. Our {\sc Cloudy} model consisted of a 1D spherical cloud geometry representing the NLR with ISM abundances, including dust grains exposed to an ionising AGN SED. The Hydrogen density at the face of the cloud was set consistent with the values from \cite{2006A&A...459...55B}, and the thickness of the cloud (used as the stopping criterion) being the same as the NLR considered in our simulations (1000\,pc). A powerlaw density profile with distance was also used with an index of -1.46 (again found when averaging the \citealp{2006A&A...459...55B} sources). The ionising AGN SED that we considered is from \cite{2012MNRAS.420.1848D}, which models the intrinsic disk emission from the optical to X-ray band from an AGN. This is done by splitting up the accretion disk into an outer disk which emits as a colour temperature corrected blackbody; a warm, optically thick inner disk of free electrons responsible for the soft excess, and a hot, optically thin corona responsible for the power-law extending up to hundreds of keV. This model has been well tested in fitting disk emission from AGN \citep{2012MNRAS.420.1848D}. Thus, this model is optimal as an input SED into {\sc Cloudy} to study the number of free electrons in the NLR. The values chosen as input parameters for the SED are the from the best fit values from a sample of 51 AGN from \cite{2012MNRAS.420.1825J} and \cite{2012MNRAS.422.3268J} (see Table\,3 in \citealp{2012MNRAS.420.1848D}). The parameters chosen for our model are a black hole mass $\log{(M_{\rm BH}/M_{\odot})} = 7.5$, an Eddington ratio $\log{(L/L_{\text{Edd}})} = -1.5$ \citep{2017ApJ...850...74K}, a temperature of the warm electrons in the inner disk $kT_{e} = 0.31\,$keV, the power-law of the hot electron corona $\Gamma = 1.87$, the optical depth of the warm electrons $\tau = 13$, and the percent of the radiation emitted in the power-law $f_{pl} = 0.34$. From these simulations we found that the column density of ionised Hydrogen, i.e. the largest contributor to the free electron column density, was $N_{\text{H}}^{\text{NLR}} = 10^{22.38}\text{\,cm}^{-2}$.


\bsp	
\label{lastpage}
\end{document}